\documentclass[11pt]{article}
\usepackage{amsmath}
\usepackage{mathrsfs}
\usepackage{amssymb}
\usepackage{titling}
\usepackage{breqn}
\usepackage{graphicx}
\usepackage{makeidx}
\usepackage{lipsum}
\allowdisplaybreaks
\usepackage{mathtools}
\usepackage[left=2cm,right=2cm,top=2cm,bottom=2cm]{geometry}
\usepackage{amsfonts}
\usepackage{authblk}
\usepackage[ansinew]{inputenc}
\usepackage[usenames, dvipsnames]{pstricks}
\usepackage{epsfig}
\usepackage{float}
\usepackage{pst-grad} % For gradients
\usepackage{pst-plot} % For axes
\usepackage{amsmath}
\usepackage{graphicx}
\usepackage{subcaption}
\usepackage{float}
\usepackage{afterpage}
\usepackage[section]{placeins}

\usepackage[font=small,labelfont=bf]{caption}

\usepackage{cite}
\DeclareRobustCommand{\rchi}{{\mathpalette\irchi\relax}}
\newcommand{\irchi}[2]{\raisebox{\depth}{$#1\chi$}} % inner command, used by \rchi
\usepackage[colorlinks, hyperindex]{hyperref}

\newcommand\scalemath[2]{\scalebox{#1}{\mbox{\ensuremath{\displaystyle #2}}}}

\hypersetup
{
colorlinks, %
citecolor=blue, %
linkcolor=Green, %
urlcolor=Cyan, %
}

\providecommand{\keywords}[1]
{	
 \textbf{\textit{Keywords---}} #1
}

\title{{{\bf Evolution of Gravitational Waves in Non-minimal Coupling Between Geometry and Matter Theories of Gravity}}}

\date{}

\author{Tahereh Azizi{\thanks{t.azizi@umz.ac.ir}}}
\author{Najibe Borhani{\thanks{najibeh.borhani@gmail.com}}}
\author{Mojtaba Haghshenas{\thanks{mjtb.haghshenas@gmail.com}}}

\affil{Department of Theoretical Physics, Faculty of Basic Sciences, University of Mazandaran,\\
	P. O. Box 47416-95447, Babolsar, IRAN }

\begin{document}

\maketitle

\begin{abstract}
We consider some specific models of non-minimal matter-geometry coupling theories and investigate the propagation of the gravitational waves in them. Extracting the temporal evolution of the gravitational wave equation within the framework of a flat FRW universe with a perfect fluid distribution, we analyze the waveforms traveling during the time. We find that while both the amplitude and frequency of the GWs decay with time in all considered models, the rate of reduction is highly sensitive to the values of the equation of state parameter and input parameters of the considered models.

\end{abstract}


\section{Introduction}
Observational data demonstrate that the universe has undergone an accelerating phase in the current epoch \cite{Buchdahl:1998,Perlmuttera:1999,Bernardisl:2000,
	Hanany:2000,Riess:2011,Ade:2016,Akrami:2018,Aghaniml:2018}. Two potential casts could take the leading role in this phenomenon, a mysterious matter component with a negative pressure dubbed as dark energy \cite{Brax:2018} and the so-called dark gravity approach corresponding to the modification of the geometric sector of the field equations without altering the matter part. Various candidates have been proposed for dark energy \cite{AmendolaL:2010},  yet fail to explain why dark energy carries the majority of the universe's energy density at present. Alternatively, a new gravitational physics has been suggested \cite{Capozziellos:2003,Carrolls:2004}, in which general relativity (GR) is modified so that the need for dark energy can be excluded. One of the simplest examples of dark gravity is to introduce an arbitrary function of the Ricci scalar into the gravitational action, which can produce cosmic inflation and current cosmic acceleration and may mimic the behavior of dark matter \cite{Buchdahl:1970,Kerner:1982,Kleinert:2002,Capozzielloc:2003,
	OdintsovN:2006,Azizi:2009,Appleby:2010,Mitchell:2021}. The terms with positive powers of curvature provide the inflation in the very early universe. In contrast, the terms with negative powers describe cosmic acceleration currently consistent with astrophysical data. It has been shown that the modified gravity with inverse curvature terms may have some problems, while its Palatini formalism gives rise to fewer drawbacks \cite{Wang:2004,dVollick:2004}.

In \cite{Nojiri:2004,Allemandi:2005} a further modification of gravity has been proposed via a non-minimal coupling with matter-like Lagrangian (serving dark energy) so that by decreasing the curvature due to universe expansion, gravitational dark energy dominance over ordinary matter. Moreover, such a gravitational alternative to dark energy may give an effective phantom phase that can describe cosmic acceleration. Such a dark coupling approach gained a lot of attention, which provides the maximal generalization of the Einstein-Hilbert Lagrangian, as exemplified by the coupling between the Ricci scalar and the matter Lagrangian density \cite{Bertolamii:2007,HarkoB:2008,HarkoL:2010}. The coupling induces the motion of the massive particles to be non-geodesic and raises an extra orthogonal force to the four-velocity. Harko et al. \cite{Harko:2011kv} have considered another extension of GR, in which the gravitational Lagrangian is given by an arbitrary function of the Ricci scalar and the trace of the energy-momentum tensor. Due to the coupling of matter and geometry, the cosmic speed-up in this modified theory results from the geometrical contribution and the matter content. A lot of research has been carried out into astrophysical and cosmological aspects of this theory, such as thermodynamics, wormhole solutions, gravitational waves, energy conditions, and so on \cite{Jamil:2012,Houndjo:2013,SharifRM:2013,Azizi:2013,SinghS:2014,Baffou:2015,Alves:2016,
	ShabaniZiaie:2017,WuLi:2018,Maurya:2020,Pretel:2021,SahooP:2021}.
%The $f(R,T)$'s linear model provides a quintessence regime and supports the acceleration of our universe \cite{SahooP:2021}.

In the literature, it has been proposed other modified theories like energy-momentum powered gravity, with some particular power values of $\beta(T_{\mu\nu}T^{\mu\nu})^{\eta}$ \cite{Katirci:2014, Roshans:2016, Board:2017, Akarsu:2018, Akarsud:2019}, where the special case $\eta=1$ is dubbed energy-momentum squared gravity. As $\eta$ determines the order of non-linearity of the energy-momentum tensor contributes to the matter Lagrangian, this model can be studied at very different energy density scales. For example, the case $\eta>\frac{1}{2}$, since the energy-momentum powered term manifests itself at larger values of $\rho$, may be effective at high energy densities that are relevant to early times. In the case $\eta<\frac{1}{2}$, the energy-momentum powered term manifests itself at lower values of $\rho$ like the present time \cite{Akarsu:2018}, it may drive Cardassian-like accelerated expansion of the universe without invoking a cosmological constant or hypothetical fluid, while for energy-momentum squared gravity at late times to drive accelerated expansion, some non-zero cosmological constant is needed.

In all the above theories, gravity is induced by curvature in Riemannian geometry, which
consists of setting the torsionless Levi-Civita connection and varying the action with
respect to the metric. To remedy some incompatibility of these theories,
it is desirable to investigate extended theories of gravity on more general geometric
structures. There is a type of space-time, the so-called Weitzenb\"{o}ck, which Einstein applied to unify the teleparallel theory of electromagnetism and gravity
\cite{Weitzenbock:1923,Einstein:1928}. Assuming a non-minimal coupling between matter and torsional geometry, 
such theories have extended to non-minimal teleparallel gravity \cite{Tharko:2014,HKOftt:2014,Carloni:2016,HKOnm:2021,Azizit:2017}.
In Weitzenb\"{o}ck space, gravity is driven
by the torsion tensor due to constraints $R^{\alpha}_{\beta\mu\nu}=0$ and $\nabla_{\alpha}g_{\mu\nu}=0$.
In this regard, another extension of Riemannian geometry is to impose cancellation of curvature and
torsion ($R^{\alpha}_{\beta\mu\nu}=0,T^{\alpha}_{\beta\mu}=0$), but instead
a non-vanishing non-metricity ($Q_{\alpha\mu\nu}=\nabla_{\alpha}g_{\mu\nu}$) is
responsible for all forms of gravitational interactions. The class of theories
based on these different underlying geometries is known as symmetric teleparallel
gravity (STG). In both metric teleparallel and symmetric teleparallel, equivalent to the ordinary formulation of general relativity can be reformulated
\cite{Ferraro:2007,Maluf:2013,Nester:1999,Jimenez:2017}.
A generalization of STG was considered in \cite{Harko:2018PRD}, by coupling between
the non-metricity $Q$ and the matter Lagrangian which can potentially explain
the accelerated expansion of the universe. In this regard, Xu et al. in \cite{HarkoLiang:2019}
introduced a more general extension of the theory wherein gravitational action
is determined by an arbitrary function of the non-metricity scalar, $Q$, and the
trace of the energy-momentum tensor, $T$. The cosmological parameters were obtained
in some functional forms of $f(Q,T)$, which favored an accelerated universe expansion using geometrical modification instead of a cosmological constant.
Observational constraints of $f(Q,T)$ gravity\cite{Aroras:2020}, cosmological implications of its
Weyl-type gravity \cite{XuHa:2020,Gadbaila:2021,Gadbails:2021}, the various energy
conditions\cite{Arorap:2020,Aroraa:2021,Aroraj:2021}, transit cosmological models of
$f(Q,T)$ gravity\cite{Pradhan:2021}, dynamical aspects and cosmic acceleration in
$f(Q,T)$ gravity \cite{PatiB:2021,Agrawal:2021,Arorap:2022,Pati:2022} and cosmological inflation
in $f(Q, T)$ gravity \cite{Shiravand:2022} shows that $f(Q,T)$ gravity can provide a consistent solution to the dark energy problem. Moreover, in a recent paper, a covariant formulation of the theory is obtained that is equivalent and provides a direct comparison with GR\cite{Loo:2023oxk}.

Nevertheless, Einstein's general theory of relativity may not be  viable at the cosmological scale,
some recent observations, like imaging the black hole shadow in the $M87$ galaxy and
the center of the Milky Way by the Event horizon telescope \cite{Event Horizon:2019, Event Horizon:2022} and the discovery of gravitational waves from the merger of black
holes and neutron stars\cite{Abbott:2016,LIGO:2017}, has approved its most
important predictions. The capability of gravitational waves to interact with matter has
provided scientists with direct detection through laser interferometers. This motivated extensive study of
gravitational waves in modified gravity theories, particularly the non-minimal ones
to investigate the possible constraints on the number of polarization modes of
gravitational waves and their associated propagation velocities\cite{Bambapgw:2013,
	Bertolamit:2018,Hohmann:2019,Capozziellopgw:2020,Haghshenas:2020,Haghshenas:2021,Najera:2022}.

In this paper, we follow another approach applied by S. Mondal et al. \cite{Mondal:2021}, which is a comprehensive study to reveal the impact of curvature of the different cosmological backgrounds with various kinds of matter distribution on waveforms of GWs in standard Einstein gravity (see also \cite{Haghshenas:2022gka} in a similar analysis for braneworld scenarios). We consider some modified theories of gravity with non-minimal coupling between matter and geometry in flat Friedmann-Robertson-Walker
(FRW) metric, to study how different values of the equation of state parameter and the modification terms represented
as dark energy, determine the amplitude of the wave and its governance period.  It is worth noting that when we consider any generalized theories of gravity, the part of modified gravity may be formally included in the total effective energy density and pressure. In fact, it is often convenient and beneficial to express the field equations of alternative theories of gravity in terms of the Einstein field equations, incorporating an effective energy-momentum tensor that encapsulates all the modifications associated with the new theory and the energy-momentum tensor of matter. As a result, the background dynamics is still GR, and we can perform the analysis for this background. Hence, in this paper, we introduce the effective energy-momentum tensor to the field equations of each model and solve the equations to obtain the scale factor in the considered scenarios.
This work is organized as follows. In section \ref{sec2}, we present
the gravitational wave equations in flat FRW space-time. In section \ref{sec3}, we briefly review some of the non-minimally matter-gravity coupling theories to obtain
the scale factor evolution of each model. We study the behavior of GWs in various matter contributions of the universe in section \ref{sec4}. Solving the gravitational equation in the presented models, we can gain further insight into the impression of additional terms in field equations due to a coupling between matter and gravity on the evolution of gravitational waves. Finally, in Section \ref{conclusion}, we present our conclusions.
\section{Gravitational wave equations in FRW background} \label{sec2}
To obtain the gravitational wave equation in the FRW universe,
we consider a general perturbed space-time metric, which deviates only slightly from a curved background metric $g_{\mu\nu}$, as follows \cite{Padmanabhan:2010zzb,Svitek:2006yj,Ford:1977dj,Flanagan:2005}
\begin{equation}\label{background}
	g^{\prime}_{\mu \nu}= g_{\mu \nu}+\epsilon h_{\mu \nu},
\end{equation}
where $\epsilon$ is a small parameter.
%As in linearized gravity, indices are raised and lowered using the metric $g_{\mu\nu}$, so the inverse of the general metric is given by
%\begin{equation}\label{backgroundinverse}
%g^{\prime\mu \nu}= g^{\mu \nu}-\epsilon h^{\mu \nu}.
%\end{equation}
%The components of the affine connection in linear order are then given by
%\begin{equation}\label{affine}
%\Gamma^\rho_{\mu \nu}=\overline{\Gamma}^\rho_{\mu\nu}+\frac{1}{2}\gamma^{\rho \sigma}[\nabla_\nu h_{\mu \sigma}+\nabla_\mu
%h_{\sigma \nu}-\nabla_\sigma h_{\mu\nu }]=\overline{\Gamma}^\rho_{\mu\nu}+\Gamma^{\rho (1)}_{\mu \nu},
%\end{equation}
%which is composed of two parts,the first term $\overline{\Gamma}$ as background space-time
%connection and the latter $\Gamma^{(1)}$ is linearly perturbed connection.
To the first order of $\epsilon$, the Riemann tensor is obtained as
\begin{equation}\label{riemann}
	{R^\rho}_{ \mu \nu \lambda}= {\overline{R}^\rho}_{ \mu \nu
		\lambda}+{R^{(1)\rho}}_{\mu \nu \lambda},
\end{equation}
where $\overline{R}^{\rho}_{\mu\nu\lambda}$ and  $R^{(1)\rho}_{\mu\nu\lambda}$ are the background and linearly perturbed curvature tensors, respectively. Consequently, the linearly perturbed part of the Ricci tensor is given by
\begin{equation}\label{perturbedpart}
	R^{(1)}_{\alpha \beta}=
	\frac{1}{2}\left[\nabla_\rho \nabla_\alpha h^\rho_\beta  +\nabla_\rho
	\nabla_\beta h^\rho_\alpha -\Box h_{\alpha\beta}-\nabla_\beta \nabla_\alpha h
	\right],
\end{equation}
where $\Box=g^{\mu\nu}\nabla_\mu \nabla_\nu$ is the wave operator and $h=h^\mu_\mu$ is the trace of the metric perturbation.
For simplification, we use the trace-reversed perturbation $\tilde{h}_{\mu \nu}= h_{\mu \nu}-\frac{1}{2} g_{\mu \nu} h$
instead of the metric perturbation $h_{\mu\nu}$. The trace-reversed condition implies that $\tilde{h}^{\mu}_{\mu}= -h$, so
one can replace $h_{\mu \nu}$ with $\tilde{h}_{\mu \nu}-\frac{1}{2} g_{\mu \nu} \tilde{h}$. We assume that the perturbation of the energy-momentum tensor is zero, which gives rise to vanishing the
linearly perturbed part of the Ricci tensor, yielding
\begin{equation}\label{perturbedzero}
	\nabla_\rho \nabla_\alpha \tilde{h}^\rho_\beta  +\nabla_\rho
	\nabla_\beta \tilde{h}^\rho_\alpha -\Box \tilde{h}_{\alpha\beta}+\frac{1}{2}g_{\mu\nu}\Box \tilde{h}=0.
\end{equation}
One can simplify the above expression further by choosing the Lorentz gauge and assuming the gravitational wave is traceless. Thus we get
$\overline{h}=\overline{h}^\alpha_\alpha=0$. Finally, the wave equation for GWs in the curved background space-time, in terms of $h_{\mu\nu}$, is given by
\begin{equation}\label{wavequationh}
	\Box {h}_{\alpha \beta}-2\bar{R}_{\sigma \alpha \beta \rho}{h}^{\sigma \rho}-\bar{R}_{\sigma \beta}
	{h}^\sigma_\alpha-\bar{R}_{\sigma \alpha}{h}^\sigma_\beta=0.
\end{equation}
In the above equation, the matter distribution obviously affects the gravitational wave propagation by the Ricci tensor, while the contribution of the background curvature is determined by the Riemann tensor.

Now we consider a flat FRW metric as the background, which is defined as
\begin{equation}\label{FRW}
	ds^2 = -dt^2 +a^2(t)dx^{i}dx^{j} ,
\end{equation}
where $a(t)$ is the scale factor that describes the dynamics of the background space-time. As we have mentioned, to investigate how the gravitational waves are evaluated in the FRW universe, first, it is required to calculate the relevant components of the Riemann and Ricci tensors. By using the equations (\ref{wavequationh}) and (\ref{FRW}), the gravitational wave equations in flat FRW, in terms of scale factor function, take the following form

\begin{eqnarray}\label{allwavequations}
	%\begin{center}
	&\Box {h}_{00} -
	2\left[\frac{\ddot{a}}{a^3}{h}_{11}+\frac{\ddot{a}}{a^3}{h}_{22}
	+\frac{\ddot{a}}{a^3}{h}_{33}\right]-6\frac{\ddot{a}}{a}{h}_{00
	} = 0,\\ \label{eq1}
	&\Box {h}_{11} +
	2\left[\frac{\dot{a}^2}{a^2}{h}_{22}+\frac{\dot{a}^2}{
		a^2}{h}_{33}-a\ddot{a}{h}_{00}\right]
	-2\left[
	2\frac{\dot{a}^2}{a^2}+\frac{\ddot{a}}{a}\right]{h}_{11}=0,\\
	\label{eq2}
	&\Box{h}_{22}+
	2\left[\frac{\dot{a}^2}{a^2}{h}_{11}+\frac{\dot{a}^2}{
		a^2}{h}_{33} - a \ddot{a}{h}_{00}
	\right]-2\left[2\frac{\dot{a}^2}{a^2}+\frac{\ddot{a}}{a}\right]{h
	}_{22} = 0,\\ \label{eq3}
	&\Box {h}_{33}- 2\left[a\ddot{a}{h}_{00}
	-\frac{\dot{a}^2}{a^2}{h}_{11}-\frac{
		\dot{a}^2}{a^2}{h}_{22}\right]-2\left[2\frac{\dot{a}^2}{a^2}+\frac{\ddot{a}}{a}\right]{h
	}_{33} = 0, \\ \label{eq4}
	&\Box {h}_{0i}-
	\left[6\frac{\ddot{a}}{a}+2\frac{\dot{a}^2}{a^2}\right]{h}_{0i}
	= 0, \text{ for}\ i = 1,2,3 ,\\ \label{eq5}
	&\Box
	{h}_{mn}-\left[2\frac{\ddot{a}}{a}+6\frac{\dot{a}^2}{a^2}\right]{
		h}_{mn} = 0, \text{ for}\ m,n = 1,2,3 \text{ and}\ m \neq n. \label{eq8}
	%\end{center}
\end{eqnarray}
In order to make the most of the transverse-traceless-synchronous (TTS) gauge, it is more convenient to use the Cartesian coordinate system ($x^{1}, x^{2}, x^{3}$ corresponding to the coordinates $x, y, z$ respectively) \cite{Grishchuk:1981bt,Caldwell:1993xw}. In this case, if we assume that the wave is propagating along the $x$ direction,
$h_{0\mu}$ and ${h}_{1\mu}$ would be zero, and ${h}_{22} = -{h}_{33}$, where the
subscript $0,1,2,3$ denotes $t,x,y,z$ coordinates respectively.
Therefore, Eq. ({\ref{allwavequations}}) reduces to the two following equations

\begin{eqnarray}
	\Box {h}_{22}-\left[2\frac{\ddot{a}}{a}+6\frac{\dot{a}^2}{a^2}\right]{h}_{22}
	&=& 0 \label{FLATFRW1}\\
	\Box {h}_{23}-\left[2\frac{\ddot{a}}{a}+6\frac{\dot{a}^2}{a^2}\right]{h}_{23}
	&=& 0\label{FLATFRW2}.
\end{eqnarray}
It is clear that the equations (\ref{FLATFRW1}) and (\ref{FLATFRW2}) are in the same form, so we will use one of them. Now, we consider an ansatz
for ${h}_{22}$ as a dual function of $t$ and $x$, which can be defined as ${h}_{22} =\Psi(x,t)$.
So the equation (\ref{FLATFRW1}) takes the following form
\begin{equation}\label{Gwaeveq}
	\frac{\partial^2 \Psi}{\partial t^2} + 3\frac{\dot{a}}{a}\frac{\partial
		\Psi}{\partial t} - \frac{1}{a^2}\frac{\partial^2 \Psi}{\partial x^2} -
	\left(2\frac{\ddot{a}}{a}+6\frac{\dot{a}^2}{a^2}\right)\Psi = 0.
\end{equation}
Since we need to know the evolution of GWs over time, we should solve the temporal part of the equation. Thereby, by using the separation of variables as $\Psi(x,t) = \chi(x)\mathcal{T}(t)$,  the equation (\ref{Gwaeveq}) is decomposed into temporal and spatial parts as follows
\begin{equation}\label{sepwave}
	\frac{\mathcal{T}(t)}{a^2}\left[\frac{d^2 \mathcal{T}(t)}{{dt}^2} +3\frac{\dot{a}}{a}\frac{d\mathcal{T}(t)}{dt} -\left(6\frac{\dot{a}^2}{a^2}+2\frac{\ddot{a}}{a}\right)\mathcal{T}(t)\right]=
	\frac{1}{\chi}\frac{d^2 \chi}{d x^2} = -{\Omega^2} ~,
\end{equation} \label{spatialpart}
where ${\Omega^2}$ is the separation constant. The solution to the spatial part is
\begin{equation}
	\chi(x) = A \cos(\Omega x) + B \sin(\Omega x),
\end{equation}
where $A$ and $B$ are integration constants. Finally, we get a differential equation whose solution is the time evolution of GWs, which is obtained by
\begin{equation}\label{temporalpart}
	\frac{d^2 \mathcal{T}(t)}{{dt}^2} +3\frac{\dot{a}}{a}\frac{d\mathcal{T}(t)}{dt} +\left(\frac{{\Omega^2}}{a^2}-6\frac{\dot{a}^2}{a^2}-2\frac{\ddot{a}}{a}\right)\mathcal{T}(t)=0.
\end{equation}

\section{Brief Review of Non-minimally Coupled Gravity Theories} \label{sec3}
In this section, we review some non-minimally coupled gravity models to get an exact solution of the scale factor in these models. It was demonstrated that, in these theories, the covariant divergence of the energy-momentum tensor is non-zero, so an energy transfer exists between matter and geometry. Indeed, such coupling between matter and geometry induces extra acceleration on the particles.

%%%%%%%%%%%%%%%%%%%%%%%%%%%%%%%%%%%%%%%%%%%%%%%%%%%%%%%%%%
\subsection{Non-minimally Coupled Scalar Theory}
%We start with the Einstein-Hilbert action for general relativity,
%\begin{equation}
%S=\int d^{4}x \sqrt{-g} \ \frac{R}{2\kappa^{2}},
%\end{equation}
%where $\kappa^{2}\equiv 8\pi G/c^{4}$.
%%In this paper we consider the Einstein gravitational constant as $\kappa^{2} = 1$ and $G=c = 1$.
%By replacing the Ricci scalar, $R$, in the above action with a generalized function of R, we get
%the action for $f(R)$ gravity as follow
%\begin{equation}
%S=\int d^{4}x \sqrt{-g} \ \frac{f\left(R\right)}{2\kappa^{2}},
%\end{equation}
To bring to light the current dark energy dominance, a consistent modified gravity model has been introduced 
in \cite{Nojiri:2004,Allemandi:2005}. Such gravitational dark energy grows because of decreasing of 
the curvature contribution to the expansion of the universe.
Let us now consider the following action as an example of such model
\begin{equation}\label{actionNMCS}
	S=\int d^4 x \sqrt{-g}\left\{ \frac{R}{\kappa^{2}}
	+ f \left(R\right) \mathcal{L}_{d}\right\}\,,
\end{equation}
where $\kappa^{2}\equiv 8\pi G/c^{4}$ and $\mathcal{L}_{d}$ is gravity-matter Lagrangian which may be interpreted as a contribution of dark energy.
We consider $f(R) = \left( \frac{R}{\xi^2} \right)^\alpha$ which $\xi$ is the coupling parameter and $\alpha$ is a constant.
Varying action (\ref{actionNMCS}) with respect to $g_{\mu\nu}$, we obtain the equation of motion as follows
\begin{equation}\label{eqmoNMCS}
	0= \frac{1}{\sqrt{-g}} \frac{\delta S}{\delta g_{\mu\nu}}
	=\frac{1}{\kappa^{2}} \left\{ \frac{1}{2}g^{\mu\nu}R - R^{\mu\nu}\right\}
	+ T^{\mu\nu\,(eff)}\,.
\end{equation}
Here $T^{\mu\nu\,(eff)}$ is the effective energy-momentum tensor that is defined in the following form
\begin{equation}\label{EMTNMCS}
	T^{\mu\nu\,(eff)} \equiv \frac{1}{\xi^{2\alpha}}\left[-\alpha R^{\alpha-1} R^{\mu\nu}\mathcal{L}_{d}+
	\alpha(\nabla^{\mu}\nabla^{\nu}-g^{\mu\nu}\nabla^{2})(R^{\alpha-1}\mathcal{L}_{d})+ R^{\alpha}T^{\mu\nu}\right]
\end{equation}
where
\begin{equation}
	T^{\mu\nu}\equiv \frac{1}{\sqrt{-g}}\frac{\delta}{\delta g_{\mu\nu}}\left(\int d^4x\sqrt{-g} \mathcal{L}_{d}\right)\,.
\end{equation}

Now we consider $\mathcal{L}_{d}$ to be the Lagrangian of the free massless scalar $L_d$ as follows
\begin{equation} \label{Lad}
	\mathcal{L}_{d} = - \frac{1}{2}g^{\mu\nu}\partial_\mu \phi \partial_\nu \phi\, .
\end{equation}
The scalar field equation is given by the variation over $\phi$, which has the following form
\begin{equation}\label{SFEQ}
	0 = \frac{1}{\sqrt{-g}}\frac{\delta S}{\delta \phi}= \frac{1}{\sqrt{-g}}
	\partial_\mu \left( \left( \frac{R}{\xi^2} \right)^\alpha \sqrt{-g} g^{\mu\nu}\partial_\nu \phi \right)\, .
\end{equation}
Considering the flat FRW metric, so the scalar $\phi$ only depends on time, leads to the solution of the field equation in the following form
\begin{equation}\label{scalarfield}
	\dot{\phi}= b a^{-3} R^{-\alpha}\,,
\end{equation}
where $b$ is an integration constant. Note that due to a non-linear coupling between matter-like Lagrangian and gravity, the second term in (\ref{actionNMCS}) can dominate when the curvature is small (large) if $\alpha>-1$ $\left(\alpha<-1\right)$.
By substitution (\ref{scalarfield}) into (\ref{eqmoNMCS}), the equation of motion given by the variation over
$g_{00}$ component takes the following form
\begin{dmath}\label{hpNMCS}
	\frac{3 H^{2}}{\kappa^{2}}-\frac{36 b^{2}}{\xi^{2 \alpha} a^{6} \left(6 \dot{H}+12 H^{2}\right)^{\alpha +2}}\left[\frac{\alpha \left(\alpha +1\right)}{4}\ddot{H}H+
	\frac{\left(\alpha +1\right)}{4}\dot{H}^{2}+\left(1+\frac{13}{4} \alpha +\alpha^{2}\right)\dot{H}H^{2}+\left(1+\frac{7 \alpha}{2}\right) H^{2}\right]= 0\,.
\end{dmath}
Here $H=\frac{\dot{a}}{a}$ is the Hubble parameter, which dot denotes a derivative with respect to the cosmic time $t$. Solving (\ref{hpNMCS}) get the explicit expression for the scale factor as a function of time as
\begin{equation}\label{scalefactorNMCS1}
	a(t) = a_0 t^{\frac{\alpha + 1}{3}} \quad \left(H= \frac{\alpha + 1}{3t}
	\right), \quad a_0^{6}\equiv\frac{\kappa^{2} b^{2}(2\alpha-1)(\alpha-1)}{3\xi^{2\alpha}(\alpha+1)^{\alpha+1}(\frac{2}{3}(2\alpha-1))^{\alpha+2}}.
\end{equation}
which implies that the deceleration parameter is $q=\frac{2-\alpha}{1+\alpha}=\frac{3}{1+\alpha}-1$.
For the matter with the equation of state $p=(\gamma-1)\rho$, where $\gamma$ is the equation of state (EoS) parameter, we have $\gamma_{\mathit{eff}}=\frac{2}{1+\alpha}$. It is worth noting that when $\alpha>2$, the universe is accelerating phase with an initial Big Bang type singularity at the origin,
which corresponds to an effective quintessence phase, i.e. $0 <\gamma_{\mathit{eff}}< \frac{2}{3}$. In the case $\alpha<-1$, to avoid a shrinking universe, we change the direction of the time as
$t \rightarrow t_{s}-t $ ($t_{s}$ is a constant), which leads to an effective phantom phase ($\gamma_{\mathit{eff}}<0$) with a future finite-time Big Rip singularity at $t=t_{s}$.
In the early universe, when the curvature is large, the Einstein term dominates for $\alpha<-1$ and $\alpha>2$.

%%%%%%%%%%%%%%%%%%%%%%%%%%%%%%%%%%%%%%%%%%%%%%%%%%%
\subsection{\boldmath{$f(R,T)$} Gravity}
Now we further modify the action for $f(R)$ gravity so that
the gravitational Lagrangian is given by $f\left(R,T\right)$ is an arbitrary function of the Ricci
scalar, $R$, and of the trace $T$ of the energy-momentum tensor of
the matter, $T_{\mu \nu}$ \cite{Harko:2011kv}.
The action takes the following form
\begin{equation}\label{actionfrt}
	S=\frac{1}{2\kappa^{2}}\int
	f\left(R,T\right)\sqrt{-g}\;d^{4}x+\int{\mathcal{L}_{m}\sqrt{-g}\;d^{4}x}\,,
\end{equation}
where $\mathcal{L}_{m}$ is the matter Lagrangian. We define the energy-momentum tensor associated to the
matter as the matter Lagrangian only depends on the metric components and not on their derivatives, then we have
\begin{equation}\label{EMT}
	T_{\mu\nu}=-\frac{2}{\sqrt{-g}}\frac{\delta(\sqrt{-g}\;\mathcal{L}_{m})}{\delta g^{\mu\nu}}=
	g_{\mu\nu}\mathcal{L}_{m}-2\frac{\partial \mathcal{L}_{m}}{\partial g^{\mu\nu}}\,.
\end{equation}
By varying the action (\ref{actionfrt}) to the metric, one
gets the field equations of $f\left( R,T\right) $  gravity
\begin{eqnarray}\label{fieldeuations}
	f_{R}\left( R,T\right) R_{\mu \nu } - \frac{1}{2}
	f\left( R,T\right) g_{\mu \nu }
	+\left( g_{\mu \nu }\square -\nabla_{\mu }\nabla _{\nu }\right)
	f_{R}\left( R,T\right)
	\nonumber \\
	=\kappa^{2} T_{\mu \nu}-f_{T}\left( R,T\right)
	T_{\mu \nu }-f_T\left( R,T\right)\Theta _{\mu \nu}\,,
\end{eqnarray}
where $f_{R}$ and $f_{T}$ denote derivates of $f\left( R,T\right) $ with respect to the $R$ and $T$, respectively and $\Theta _{\mu \nu}$is determined by

\begin{equation}\label{theta}
	\Theta_{\mu \nu}=g^{\alpha \beta }\frac{\delta T_{\alpha \beta
	}}{\delta g^{\mu \nu}}=-2T_{\mu\nu}+g_{\mu\nu}\mathcal{L}_{m}-2g^{\alpha\beta}\frac{\partial^{2}\mathcal{L}_{m}}{\partial g^{\mu\nu}\partial g^{\alpha\beta}}\,.
\end{equation}
As it can be seen from the above equation, $\Theta_{\mu\nu}$ depends on matter Lagrangian explicitly.
We consider a perfect fluid form for the energy-momentum tensor of
the matter, given by $T_{\mu \nu}=\left(\rho +p\right)U_{\mu }U_{\nu}+pg_{\mu \nu}$,
where $\rho$ and $p$ are the energy density pressure of the perfect
fluid respectively. $U_{\mu }$ is the four-velocity of the fluid which satisfies the condition
$U_{\mu} U^{\mu}=1$. Assuming the matter Lagrangian as $L_{m}=-p$, hence for the case of perfect fluid (\ref{theta}) becomes
\begin{equation}
	\Theta _{\mu \nu }=-2T_{\mu \nu }-pg_{\mu \nu }\,.
\end{equation}
We consider the $f\left( R,T\right) $ model by a simple form, i.e,  $f\left(R,T\right)=R+2f(T)$. Therefore, the gravitational field equations (\ref{fieldeuations}) take the following form
\begin{equation}\label{fieldperfect}
	R_{\mu\nu}-\frac{1}{2}Rg_{\mu\nu}=\kappa^{2}T_{\mu\nu}^{(eff)}\,,
\end{equation}
where the prime is a derivative with respect to $T$ and the effective energy-momentum tensor is defined as
\begin{equation}
T_{\mu\nu}^{(eff)}=T_{\mu\nu}
+\frac{1}{\kappa^{2}}\left[2f_{T}T_{\mu\nu}+\left(2pf_{T}+f(T)\right)g_{\mu \nu }\right]\,.
\end{equation}
By choosing $f(T)=\lambda T$ that $\lambda $ is a coupling constant, the substitution of the FRW metric (\ref{FRW})
in the field equation (\ref{fieldperfect}) yields the Friedmann equations as follows
\begin{equation}\label{Friedmannfrt1}
	3H^{2}=\left(\kappa^{2} +3\lambda \right)\rho\ - \lambda p ,
\end{equation}
\begin{equation}\label{Friedmannfrt2}
	2\dot{H}+3H^{2}= -\left(\kappa^{2} +3\lambda \right)p + \lambda \rho\,.
\end{equation}
Now we consider the content of the universe as a perfect fluid with an equation of state in the
form of $p=(\gamma-1)\rho$, which $\gamma$ is a constant known as the EoS parameter of the
perfect fluid \cite{Singh:2014,Singh:2016}, using (\ref{Friedmannfrt1}) and (\ref{Friedmannfrt2})
one gets the single following equation
\begin{equation}\label{Friedmannfrt3}
	\dot{H}+\frac{3\gamma}{2} \ \frac{\left(\kappa^{2} +2\lambda \right)}{\left(\kappa^{2} +4\lambda-\gamma\lambda \right)}\ H^{2}=0\,.
\end{equation}
Solving (\ref{Friedmannfrt3}) for $\gamma\neq0$, we obtain
\begin{equation}
	H(t)=\frac{1}{C+\frac{3\gamma}{2} \frac{\left(\kappa^{2} +2\lambda \right)}{\left(\kappa^{2} +4\lambda-\gamma\lambda \right)}t}\,,
\end{equation}
where $C$ is an integration constant. From the above equation,  we get the scale factor as follows
\begin{equation}\label{scalefactorfrt}
	a(t)=a_0 \left(1+  \frac{3}{2}\gamma_{\mathit{eff}} H_0 (t-t_{0}) \right)^{\frac{2}{3\gamma_{\mathit{eff}}}}\,,
\end{equation}
where $\gamma_{\mathit{eff}}=\frac{\gamma\left(\kappa^{2} +2\lambda \right)}{\left(\kappa^{2} +4\lambda-\gamma\lambda \right)}$
and $H_{0}=H(t_{0})$. Moreover, one can obtain the deceleration parameter as
\begin{equation}
	q=-\frac{a\ddot{a}}{\dot{a}^{2}}=\bigg[\frac{3\gamma}{2}\frac{(\kappa^{2} +2\lambda)}{(\kappa^{2} +4\lambda-\gamma \lambda)}-1\bigg]=\frac{3\gamma_{\mathit{eff}}}{2}-1\,.
\end{equation}
Obviously, to have an accelerating universe, it should be $0<\gamma_{\mathit{eff}}<\frac{2}{3}$.

%%%%%%%%%%%%%%%%%%%%%%%%%%%%%%%%%%%%%%%%%%%%%%%%%%%%%%%5
\subsection{\boldmath{$f(R,T_{\mu\nu} T^{\mu\nu})$} Gravity}
%%%%%%%%%%%%%%%%%%%%%%%%%%%%%%%%%%%%%%%%%%%%%%%%%%%%%%%%%%
Here we consider the Energy-momentum powered gravity  with the following action \cite{Katirci:2014}
\begin{equation}\label{action3}
	S=\int f(R,T_{\mu\nu}T^{\mu\nu})\sqrt{-g}\;d^4x+\int \mathcal{L}_{m}\sqrt{-g}\;d^4x .
\end{equation}
Taking a variation of the action (\ref{action3}) with respect to the metric, one can obtain the field equation as follows
\begin{equation}\label{fieldeqfrtt2}
	f_RR_{\mu\nu}-\frac{1}{2}fg_{\mu\nu}+(g_{\mu\nu}\nabla_{\alpha}\nabla^{\alpha}-\nabla_{\mu}\nabla_{\nu})f_R=\frac{1}{2}T_{\mu\nu}-f_{T^2}\Theta _{\mu \nu}\,,
\end{equation}

where we have defined $f_R=\frac{\partial f(R,T_{\mu\nu}T^{\mu\nu})}{\partial R}$ and $f_{T^2}=\frac{\partial f(R,T_{\mu\nu}T^{\mu\nu})}{\partial(T_{\mu\nu}T^{\mu\nu})}$.
and the new tensor $\Theta _{\mu \nu}$ is defined as
\begin{equation}
	\Theta _{\mu \nu}=\frac{\delta(T_{\alpha\beta}T^{\alpha\beta})}{\delta g^{\mu\nu}}=-2\mathcal{L}_{m}\left(T_{\mu\nu}-\frac{1}{2}g_{\mu\nu}T\right)-TT_{\mu\nu}+2T_{\mu}^{\alpha}T_{\nu\alpha}-4T^{\alpha\beta}\frac{\partial^2
		\mathcal{L}_{m}}{\partial g^{\mu\nu}\partial g^{\alpha\beta}}\,.
\end{equation}
Considering a perfect fluid with $\mathcal{L}_{m}=p$ , we obtain
\begin{equation}\label{thetafrtt}
	\Theta_{\mu\nu}=(-\rho^{2}-4\rho p -3p^{2})u_{\mu}u_{\nu}\,.
\end{equation}

Now we study the cosmological dynamics of the model in flat FRW background.
Assuming $f(R,T_{\mu\nu}T^{\mu\nu})=\frac{R}{2\kappa^{2}}+\beta\sqrt{T_{\mu\nu}T^{\mu\nu}}$, where $\beta$ is a coupling parameter, hence the equation (\ref{fieldeqfrtt2}) becomes
\begin{equation}\label{feqfrttmodel}
	R_{\mu\nu}-\frac{1}{2}g_{\mu\nu}R=\kappa^{2} T_{\mu\nu}^{(eff)}
\end{equation}
where we have defined the effective energy-momentum tensor as
\begin{equation}\label{feqfrttmodel}
T_{\mu\nu}^{(eff)}=T_{\mu\nu}+\beta g_{\mu\nu}\sqrt{T_{\alpha\beta}T^{\alpha\beta}}
	- \beta\frac{\Theta_{\mu\nu}}{\sqrt{T_{\alpha\beta}T^{\alpha\beta}}},
\end{equation}
Using (\ref{thetafrtt}) and (\ref{feqfrttmodel}), we obtain the Friedmann equations as follows
\begin{equation}
	3H^2=\kappa^{2}\rho~\bigg(1+\beta\frac{4(\gamma-1)}{\sqrt{1+3(\gamma-1)^2}}\bigg)=\rho_{\mathit{eff}}\,,\label{friedmanfrtt1}
\end{equation}
\begin{equation}
	2\dot H+3H^2=\kappa^{2}\rho~\bigg(1-\gamma-\beta\sqrt{1+3(\gamma-1)^2}\bigg)=p_{\mathit{eff}}\label{friedmanfrtt2}\,,
\end{equation}
where $\rho_{\mathit{eff}}$ and $p_{\mathit{eff}}$ are the effective energy and pressure, respectively. Note that the case $\beta=0$ corresponds to the GR. Plugging \eqref{friedmanfrtt1} into \eqref{friedmanfrtt2}, one can obtain
\begin{equation}\label{phubfrtt}
	\dot H=-\frac{3}{2} \gamma_{\mathit{eff}} H^2 ,
\end{equation}
where we have defined
\begin{equation}
	\gamma_{\mathit{eff}}=\frac{\gamma+\beta\frac{3 (\gamma-1)^2+4\gamma-3}{\sqrt{3 (\gamma-1)^2+1}}}{1+\beta\frac{4(\gamma-1)}{\sqrt{3 (\gamma-1)^2+1}}}\,.
\end{equation}
Using Eq. (\ref{phubfrtt}), the scale factor is obtained as follows
\begin{equation}\label{scalefactorfrtt}
	a(t)=t^{\frac{2}{3\gamma_{\mathit{eff}}}}.
\end{equation}
As the energy density varies depending on $\beta$ as $\rho=\rho_{0}(\frac{a}{a_{0}})^{-3\gamma_{\mathit{eff}}}$, for the
matter dominated universe, only the sign of $\beta$ affect on the speed of reduction of density. On the other hand, for $\gamma=0$, the energy density becomes constant, and $\gamma_{\mathit{eff}}$ is independent of $\beta$.
In the case of radiation($\gamma=\frac{4}{3}$), the sign and the magnitude of $\beta$, are essential for the evolution of the density of radiation.

\subsection{\boldmath{$f(Q,T)$} Gravity}
%%%%%%%%%%%%%%%%%%%%%%%%%%%%%%%%%%%%%%%%%%%%%%%%%%%%%%%%%%%%%%%%%
$f(Q)$ gravity is constructed in analogy with $f(R)$ gravity, where the Lagrangian
is taken to be an arbitrary function of the non-metricity scalar\cite{Jimenez:2017},
on a different underlying geometry in which curvature and torsion are both imposed equal to zero.
This novel modified gravity is a viable alternative to dark energy in addressing the problem of accelerated expansion
of the universe, which has drawn attention very recently.\\
Now we consider a more general extension of the $f(Q)$ gravity \cite{HarkoLiang:2019}, in which the non-metricity $Q$ is coupled non-minimally to the trace of the matter energy-momentum tensor via the following action
\begin{equation}\label{actionQT}
	S=\int \left[\frac{1}{2\kappa^{2}}f\left(Q,T\right) + \mathcal{L}_{m}\right]\sqrt{-g}\;d^{4}x \,
\end{equation}

where the non-metricity scalar $Q$ is defined as \cite{Hehl:1976},
\begin{equation}
	Q=-g^{\mu\nu}\left(L^{\alpha}_{\ \beta\mu}L^{\beta}_{\ \nu\sigma}-L^{\alpha}_{\ \beta\alpha}L^{\beta}_{\ \mu\nu}\right).
\end{equation}

Here $L_{~\beta \gamma}^{\alpha }$ is the disformation tensor that is given by
\begin{equation}
	L_{~\beta \gamma}^{\alpha }=-\frac{1}{2} g^{\alpha \lambda}(\nabla_{\gamma} g_{\beta\lambda}+\nabla_{\beta} g_{\lambda\gamma}-\nabla_{\lambda} g_{\beta\gamma} ).
\end{equation}
Now we introduce the non-metricity conjugate (or superpotential)\cite{Jimenez:2017}
\begin{eqnarray}
	4P^{\alpha }{}_{\mu \nu } &=& -Q^{\alpha }{}_{\mu \nu } + 2Q_{(\mu %
		\phantom{\alpha}\nu )}^{\phantom{\mu}\alpha } - Q^{\alpha }g_{\mu \nu }
	-\tilde{Q}^{\alpha }g_{\mu \nu }-\delta _{(\mu }^{\alpha }Q_{\nu )}\,, \label{super}
\end{eqnarray}
where  $Q_{\rho\mu\nu}\equiv\nabla_{\rho} g_{\mu\nu}$ is the non-metricity tensor and
$Q_{\alpha}\equiv Q_{\alpha\ \ \ \mu}^{\ \ \mu}, \tilde{Q}_{\alpha}\equiv Q^{\mu}_{\ \  \alpha \mu}$
which satisfies $Q=-Q_{\alpha \mu \nu }P^{\alpha \mu \nu}$. Taking a variation of the action (\ref{actionQT}) with respect to the metric tensor yields the following field equations as
\begin{equation}\label{fieldQT1}
	\kappa^{2}T_{\mu\nu}= -\frac{2}{\sqrt{-g}}\nabla_{\alpha} \left(f_{Q}\sqrt{-g}\ P^{\alpha}_{\ \mu\nu} \right)- \frac{1}{2}f g_{\mu\nu} \ + f_{T}\left( T_{\mu\nu} + \Theta_{\mu\nu}\right)
	-f_{Q}\left( P_{\mu\alpha\beta}\ Q_{\nu}^{\ \alpha\beta} -2 Q^{\alpha\beta}_{\ \  \mu}\ P_{\alpha\beta\nu}\right)\,,
\end{equation}
where $f_{Q}$ represents a derivative of $f$ with respect to $Q$ and $T_{\mu\nu}$ and $\Theta_{\mu\nu}$ are defined as equations (\ref{EMT}) and (\ref{theta}), respectively.

In addition, varying the action with respect to the connection, the connection field equation is obtained in the following form \cite{JimenezHeisenberg:2018}
\begin{equation}\label{fieldQT2}
	\nabla_{\mu}\nabla_{\nu}\bigg( \sqrt{-g}f_{Q} P^{\mu\nu}_{\ \ \ \ \alpha} +\frac{\kappa^{2}}{2}\pi H_{\alpha}^{\ \ \mu \nu} \bigg)=0\,,
\end{equation}
where $H_{\lambda }^{\mu \nu }$ is the hypermomentum tensor density that is defined as
\begin{equation}
	H_{\lambda }^{\mu \nu }=-\frac{1}{2}\frac{\delta (\sqrt{-g}\mathcal{L}_{m})}{\delta \Gamma _{\phantom{\lambda}\mu \nu }^{\lambda }}\,,
\end{equation}
Following \cite{Loo:2023oxk}, the field equations of $f(R,Q)$ gravity can be recast via the covariant formulation as
\begin{equation}
f_{Q}G_{\mu\nu}+\frac{1}{2}g_{\mu\nu}(Qf_{Q}-f)+f_{T}\left( T_{\mu\nu} + \Theta_{\mu\nu}\right)+
2(\nabla_{\alpha}f_{Q})P_{\mu\nu}^{\alpha}=\kappa^{2}T_{\mu\nu},
 \end{equation}
where $G_{\mu\nu}$ is the Einstein tensor corresponding to the Levi-Civita connection and the effective energy-momentum
tensor can be defined as \cite{Loo:2023oxk}
\begin{equation}
\kappa^{2}T_{\mu\nu}^{(eff)}=\kappa^{2}T_{\mu\nu}-f_{T}\left( T_{\mu\nu} + \Theta_{\mu\nu}\right)-\frac{1}{2}g_{\mu\nu}(Qf_{Q}-f)-2(f_{QQ}\nabla_{\alpha}Q+f_{QT}\nabla_{\alpha}T)P_{\mu\nu}^{\alpha}.
\end{equation}

 Assuming the perfect fluid for the matter content of the universe in flat FRW metric, we obtain the following generalized Friedmann equations
\begin{equation}\label{fieldQT3}
	\kappa^{2}\rho =\frac{f}{2}-6f_{Q}H^2 - \frac{2f_{T}}{1+f_{T}}\left(\dot{f_{Q}}H +f_{Q} \dot{H} \right),
\end{equation}
\begin{equation}\label{fieldQT4}
	\kappa^{2}p= -\frac{f}{2}+6f_{Q}H^2+ 2\left(\dot{f_{Q}}H +f_{Q} \dot{H}\right).
\end{equation}
The above equations lead to the evolution equation for the Hubble function $H$ as
\begin{equation}\label{evolutionHQT}
	\dot{H} + \frac{\dot{f_{Q}}}{f_{Q}}H =\frac{\kappa^{2}}{2f_{Q}} \left( 1+f_{T} \right)( \rho + p ).
\end{equation}
From Equations (\ref{fieldQT3}) and (\ref{evolutionHQT}), the barotropic matter density is given by
\begin{equation}\label{mdensityQT}
	\rho =\frac{f-12f_{Q}H^2}{2\kappa^{2}\left(1+\gamma f_{T}\right)}.
\end{equation}
In the following, a cosmological model of $f(Q,T)$ gravity is considered, which is constructed by
using a simple functional form of this function as $f(Q,T)=\mu Q^{n+1}+\nu T$ ,
which $\mu$, $\nu$ and $n$ are arbitrary constants.
In this case, the expression of the matter density takes the following form
\begin{equation}\label{roHQT}
	\rho =\frac{  6^{n+1} (2 n+1)\mu H^{2 (n+1)}}{ \nu (\gamma-4)-2\kappa^{2}}.
\end{equation}
Hence, the Hubble evolution equation is obtained as
\begin{equation}
	\dot{H}+\frac{3 (\nu +\kappa^{2} ) \gamma  H^2}{(n+1) (2\kappa^{2} -\nu  (\gamma-4))}=0,
\end{equation}
Consequently, the Hubble parameter is given by
\begin{equation}
	H(t)=\frac{H_{0}(n+1)\bigg( 2\kappa^{2} -\nu (\gamma -4)\bigg) }{3(\kappa^{2}+\nu
		)\gamma H_{0}\left( t-t_{0}\right) +(n+1)\bigg( 2\kappa^{2} -\nu (\gamma -4)\bigg)  }.
\end{equation}
As a result, the evolution of the scale factor is given by
\begin{equation}\label{scffqt}
	a(t)=a_{0}\bigg(a_{1}+ 3 \gamma(\kappa^{2}+\nu )  H_{0}(t-t_{0})\bigg) ^{\frac{a_{1}}{3\gamma(\kappa^{2}+\nu ) }},
\end{equation}%
where $a_0$ is an integration constant and $a_{1}$ has denoted as  $a_{1}=(n+1) \left(2\kappa^{2} -\nu  (\gamma -4)\right)$.
It has been demonstrated that, to have an accelerating expansion, a constraint imposed on the model parameters as $\frac{\gamma(\kappa^{2}+\nu )}{a_{1}}<\frac{1}{3}$ \cite{HarkoLiang:2019}.

\section{Gravitational Waves in Non-minimal Coupled Gravity Theories} \label{sec4}
In this section, we study the evolution of the waveform of GWs for the models that are considered in the previous section.
Using the scale factors evolution representing the dynamics of those models,
we obtain the exact solution of their wave equation. In order to do this, it is required to plug the scale factor of each model into the time-dependent part of the wave equation (\ref{temporalpart}).

\subsection{Evolution of the Waveform in Non-minimally Coupled Scalar Theory}
\begin{figure}[H]
	\centering
	\includegraphics[scale=0.5]{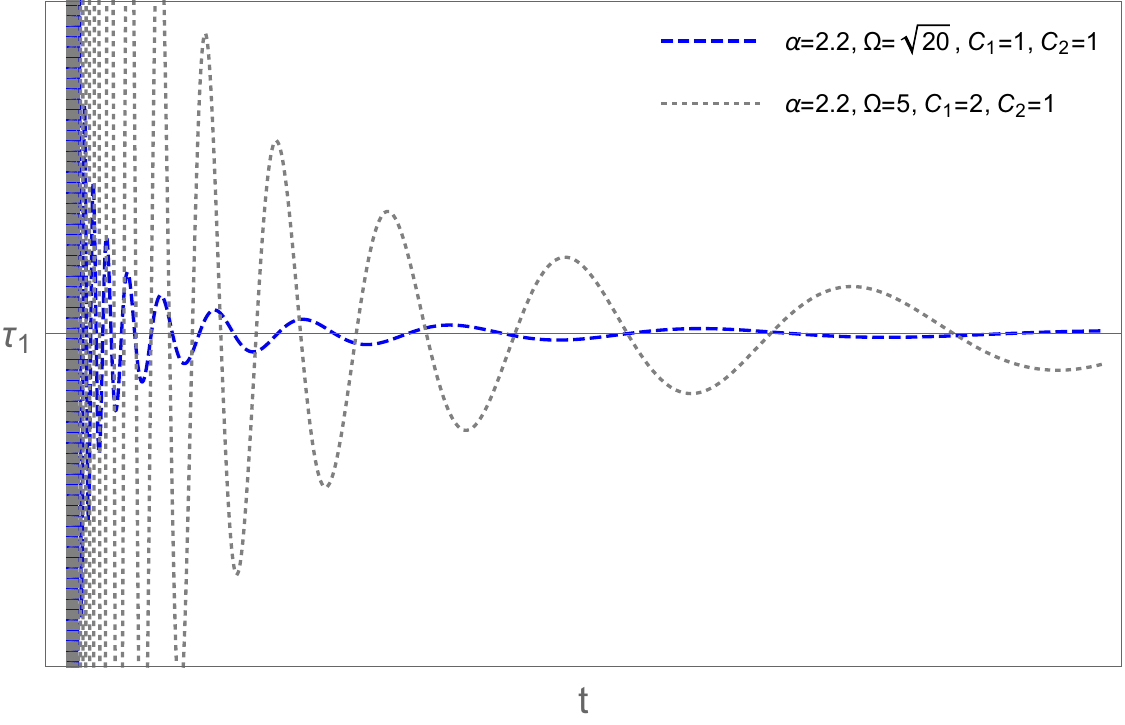}
	\caption{Evolution of gravitational wave for non-minimally coupled scalar theory
		in quintessence phase ($\alpha>2$) in flat FRW universe. (The time axis is an arbitrary unit)}
	\label{fig1}
\end{figure}

Substituting (\ref{scalefactorNMCS1}) into (\ref{temporalpart}), the temporal part of the wave equation for non-minimally coupled scalar theory takes the following form
\begin{equation}
	\frac{d^{2}\mathcal{T}_{1}}{dt^{2}}+\frac{(\alpha +1)}{t}\frac{d\mathcal{T}_{1}}{dt}+
	\left(\frac{\Omega ^2 }{a_0^2 t^{\frac{2}{3}(\alpha +1) }}-\frac{2 (4\alpha +1) (\alpha +1)}{9 t^2} \right)\mathcal{T}_1 =0\,.
\end{equation}
Indeed, the above wave equation is a Bessel differential equation which has a
general solution as sum of the first kind of order $n$ Bessel function $J_{n}(z)$
\begin{align}
	\begin{array}{ll}
		\mathcal{T}_1(t) &=  t^{-\frac{\alpha }{2}}\left(\frac{a_0 (4-2 \alpha )}{3 \Omega }\right){}^
		{\frac{3\alpha }{4-2 \alpha }}  \bigg[ C_1 \Gamma \left(1-\frac{\sqrt{41 \alpha ^2+40 \alpha +8}}
		{4-2 \alpha }\right)J_{-\frac{\sqrt{41 \alpha ^2+40 \alpha +8}}{4-2 \alpha }}\left(\frac{3 t^{\frac{2-\alpha }{3}} \Omega }{a_0(2-\alpha ) }\right) \\
		& + C_2 \ \Gamma \left(1+\frac{\sqrt{41 \alpha ^2+40 \alpha +8}}{4-2 \alpha }\right) J_{\frac{\sqrt{41 \alpha ^2+40
					\alpha +8}}{4-2 \alpha }}\left(\frac{3 t^{\frac{2-\alpha}{3}} \Omega }{ a_0(2-\alpha )}\right)  \bigg]\,,
	\end{array}
\end{align}
where $C_{1}$ and $C_{2}$ are the constants of integration and $\Gamma $ is the gamma function.
In FIG. \ref{fig1}, we plot $\mathcal{T}_1$ with respect to the time for the case $\alpha>2$, where $\kappa^{2}$, $b$ and $\xi$  are set to unity. 
%%%%%%%%%%%%%%%%%%%%%%%%%%%%%%%%%%%%%%%%%%%%%%%%%%%%%%%%%%%%%%%%%%%%%%%%%%%%%%%%%%
By comparing FIG. \ref{fig1} with the results obtained from the study of GWs propagation in different cosmological backgrounds within the framework of General Relativity (GR) \cite{Mondal:2021}, we observe a significant similarity in the behavior of GWs in this model and the dust-dominated case in GR theory. However, in the Non-Minimal Coupling theory, GWs decay to some extent, which can be attributed to the higher rate of cosmic expansion in the Non-Minimal Coupling model. As $\alpha$ increases, the scale factor increases and consequently the rate of GWs decay also increases.
%%%%%%%%%%%%%%%%%%%%%%%%%%%%%%%%%%%%%%%%%%%%%%%%%%%%%%%%%%%%%%%%%%%%%%%%%%%%%%%%%%

%It is clear that the amplitude of the wave decays, and the frequency of the wave reduces with respect to time, as well. According to the plot, one can conclude that for greater powers of the Ricci scalar, both the amplitude and frequency of the gravitational wave decrease more rapidly with time.

\begin{figure}[H]
	\centering
	\includegraphics[scale=0.6]{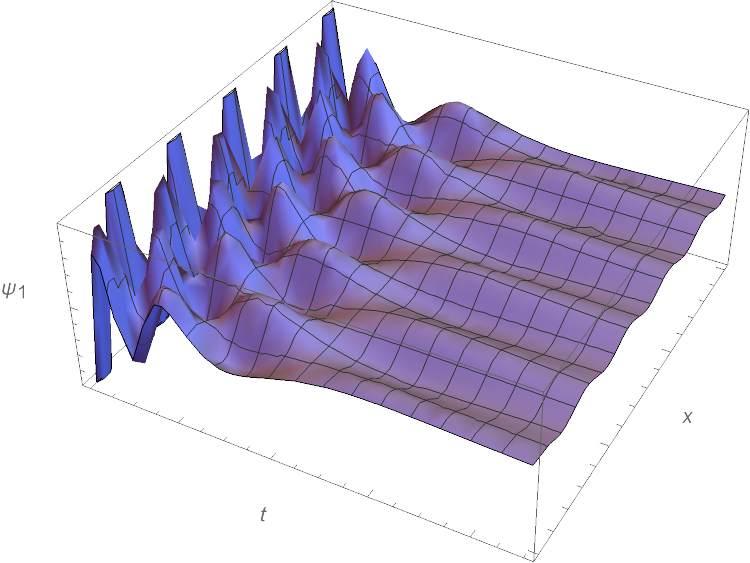}
	\caption{3D plot for the behavior of a test wave $\psi_{1}(x,t)=\rchi_{1}(x)\mathcal{T}_{1}(t)$
		in non-minimally coupled scalar theory in quintessence phase ($\alpha>2$) in flat FRW universe (t and x axes are in arbitrary units).}
	\label{fig2}
\end{figure}

\begin{figure}[H]
	\centering
	
	\centering
	\includegraphics[scale=0.43]{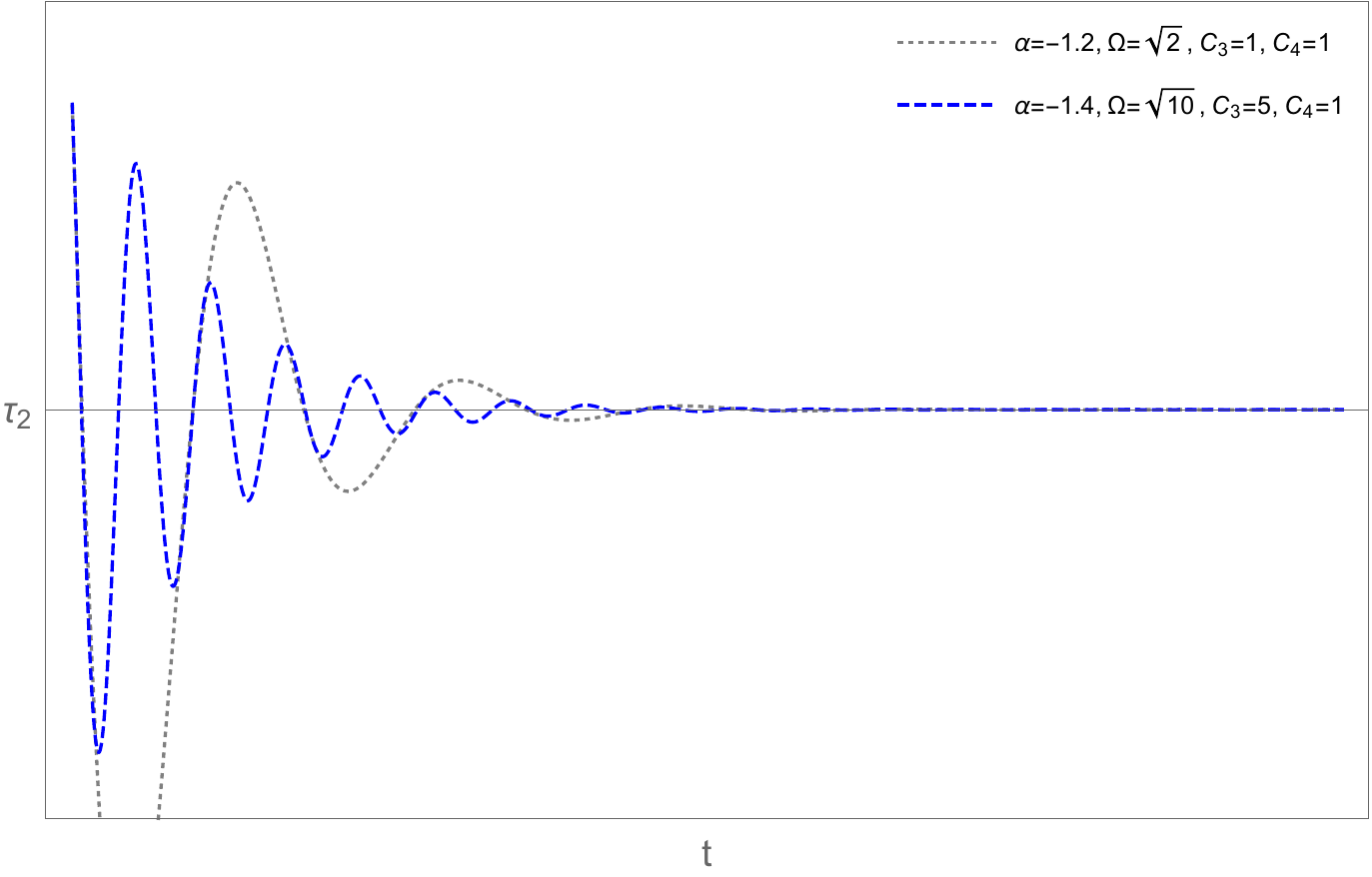}
	\caption{Evolution of gravitational wave in non-minimally coupled scalar theory for
		quintessence regime ($\alpha<-1$) flat FRW universe. (The time axis is an arbitrary unit)}
	\label{fig3}
\end{figure}

As we have mentioned earlier, for $\alpha<-1$ the scale factor (\ref{scalefactorNMCS1}) is recast as $a(t) = a_0 (t_{s}-t)^{\frac{\alpha + 1}{3}}$. Consequently, in this case, the temporal part of the wave equation reads as
\begin{equation}
	\frac{d^{2}\mathcal{T}_{2}}{d t^{2}}-\frac{\alpha +1}{t_s-t}\frac{d\mathcal{T}_{2}}{dt}+
	\left[\frac{\Omega ^2}{a_0^2 \left(t_s-t\right){}^{\frac{2 }{3}(\alpha +1)}}-\frac{2 (4 \alpha +1)
		(\alpha +1) }{9 \left(t_s-t\right){}^2} \right]\mathcal{T}_{2}=0\,.
\end{equation}
The above differential equation has a solution as follows
\begin{align}
	\begin{array}{ll}
		\mathcal{T}_2(t) &=  C_3\left(\frac{3 \Omega }{a_0(2 \alpha -4) }\right){}^{-\frac{\sqrt{41
					\alpha ^2+40 \alpha +8}}{2 \alpha -4}} \left(t_s-t\right){}^{ -\frac{\alpha }{2}-\frac{\sqrt{41
					\alpha ^2+40 \alpha +8}}{3}}\Gamma \left(1+\frac{\sqrt{41 \alpha ^2+40 \alpha +8}}{2 \alpha -4}
		\right)J_{\frac{\sqrt{41 \alpha ^2+40 \alpha +8}}{2 \alpha -4}}\left(\frac{3 \Omega  \left(t_s-t\right){}^
			{\frac{2-\alpha }{3}}}{a_0(\alpha -2) }\right) \\
		& + C_4 \left(\frac{3 \Omega }{a_0(2 \alpha -4) }\right){}^{\frac{\sqrt{41 \alpha ^2+40 \alpha +8}}{2 \alpha -4}}
		\left(t_s-t\right){}^{-\frac{\alpha }{2}+\frac{\sqrt{41 \alpha ^2+40 \alpha +8}}{3} }\Gamma \left(1-
		\frac{\sqrt{41 \alpha ^2+40 \alpha +8}}{2 \alpha -4}\right)J_{-\frac{\sqrt{41 \alpha ^2+40 \alpha +8}}
			{2 \alpha -4}}\left(\frac{3 \Omega  \left(t_s-t\right){}^{\frac{2-\alpha }{3}}}{a_0(\alpha -2) }\right)
	\end{array}\,,
\end{align}
where $C_3$ and $C_4$ are integration constants. The behavior of $\mathcal{T}_2$ versus time is depicted in FIG. \ref{fig3}.
%%%%%%%%%%%%%%%%%%%%%%%%%%%%%%%%%%%%%%%%%%%%%%%%%%%%%
For $\alpha<-1$, because of the effective phantom phase, which leads to the fate of a Big Rip for the universe, the rate of GWs decay increases significantly (see FIG. \ref{fig3}), and it is consistent with the fact that the universe undergoes perpetual expansion in this model.
It is worth noting that based on the similarities in the behavior of GWs between this model and GR theory, it can be inferred that as this model can successfully describe the accelerating rate of cosmic expansion, it can also be helpful in uncovering the behavior of GWs by appropriately selecting $\alpha$ and constants.

%%%%%%%%%%%%%%%%%%%%%%%%%%%%%%%%%%%%%%%%%%%%%%%%%%%%%%%%%%%%%%%%%55
%The figure shows that, similar to positive $\alpha$, the amplitude and
%frequency of the wave decrease with time for $\alpha<-1$, but the rate
%of decaying is significantly faster compared with the previous case.
%As a result, more negative powers of the curvature cause the amplitude and frequency of the wave to decrease faster.

\begin{figure}[H]
	\centering
	\includegraphics[scale=0.6]{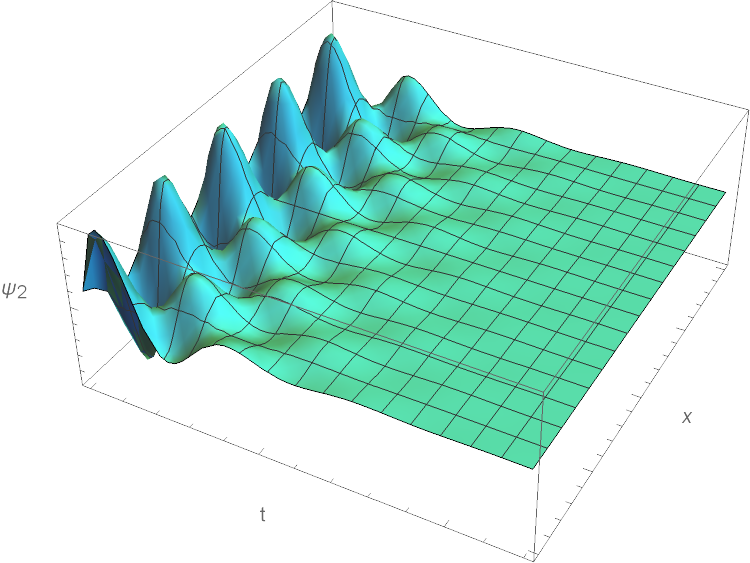}
	\caption{3D plot for the behavior of a test wave $\psi_{2}(x,t)=\rchi_{2}(x)\mathcal{T}_{2}(t)$
		in non-minimally coupled scalar theory for phantom regime ($\alpha<-1$) in flat FRW universe. (t and x axes are arbitrary units)}
	\label{fig4}
\end{figure}

%%%%%%%%%%%%%%%%%%%%%%%%%%%%%%%%%%%%%%%%%%%%%%%%%%%%%%%%%%%%%%%%%%%%%%%%%%%%%%%%%%%%%%%%%%%%%%%%%%%%%%%%%
\subsection{Evolution of the Waveform in \boldmath{$f(R,T)$} Gravity}
\begin{figure}[H]
	\centering
	\includegraphics[scale=0.52]{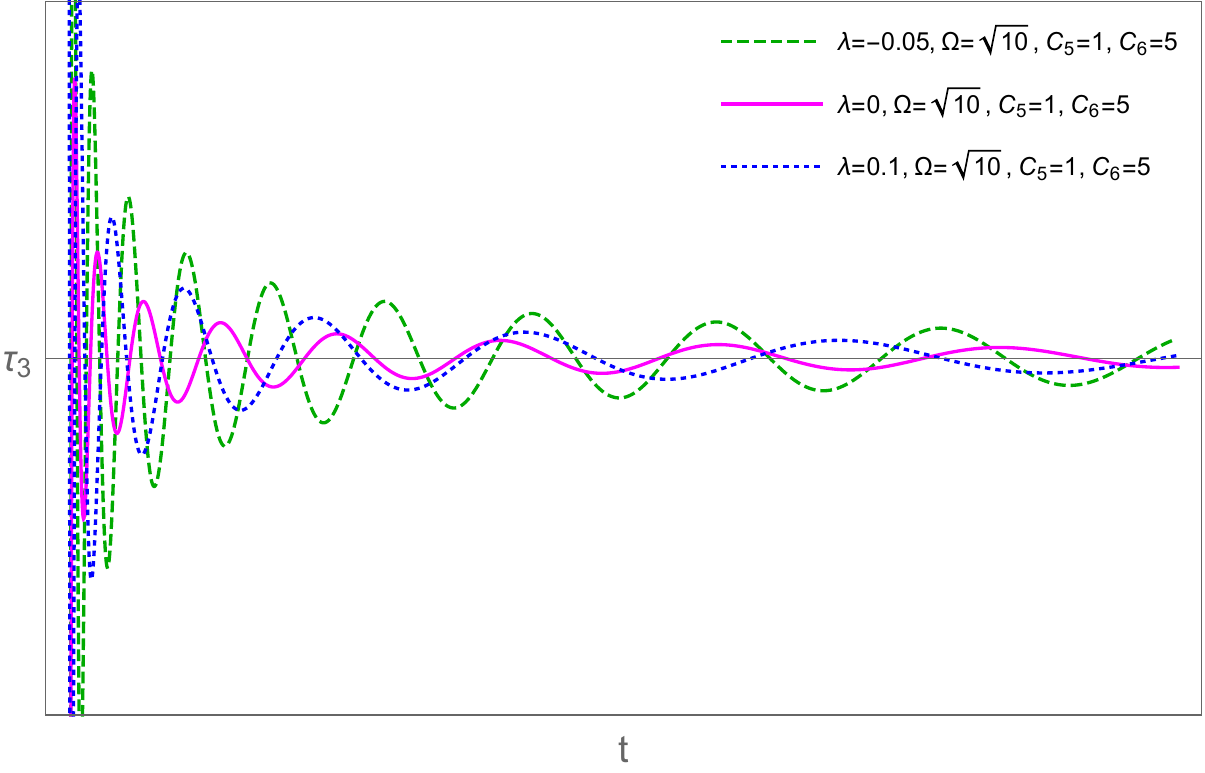}
	\caption{Evolution of gravitational wave for $f(R,T)$ gravity theory in
		dust-dominated flat FRW universe ($\gamma=1$). (The time axis is an arbitrary unit)}
	\label{fig5}
\end{figure}

In this subsection, we investigate the propagation of GWs in the $f(R,T)$ model, which we have discussed in the previous section.
Using the equations (\ref{temporalpart}) and (\ref{scalefactorfrt}), one can obtain the temporal evolution of the wave equation as follows
\begin{center}
$$\scalemath{0.95}{\frac{d^{2}\mathcal{T}_{3}}{dt^{2}}+\frac{6 H_0}{2+ 3 \gamma_{\mathit{eff}}  H_0 \left(t-t_0\right)}\frac{d\mathcal{T}_{3}}{dt}+\left[\frac{\Omega ^2 }{a_0^2 \ \bigg(1+ \frac{3}{2} \gamma_{\mathit{eff}}  H_0 \left(t-t_0\right)\bigg){}^{\frac{4}{3 \gamma_{\mathit{eff}} }}}+\frac{4\ H_0^2 \left(3 \gamma_{\mathit{eff}} -8\right)}{\bigg(2+ 3 \gamma_{\mathit{eff}}  H_0 \left(t-t_0\right)\bigg){}^2}\right]\mathcal{T}_{3} =0}$$.
\end{center}
%here $\gamma_{\mathit{eff}}=\frac{\gamma\left(1 +2\lambda \right)}{\left(1 + 4\lambda-\gamma\lambda \right)}$.
The solution of the above differential equation read as
\begin{align}
	\begin{array}{ll}
		\mathcal{T}_{3}(t) & = \frac{\left(1+\frac{3}{2} H_0 \left(t-t_0\right) \gamma _{\mathit{eff}}\right){}^
			{\frac{1}{\gamma _{\mathit{eff}}}+\frac{1}{2}}}{2^{\frac{1}{\gamma _{\mathit{eff}}}}}
		\bigg[C_5 \left(\frac{\Omega }{a_0 H_0 \left(3 \gamma _{\mathit{eff}}-2\right)}\right){}^
		{-\frac{\sqrt{9 \gamma _{\mathit{eff}}^2-84 \gamma _{\mathit{eff}}+164}}{6 \gamma _{\mathit{eff}}-4}}
		\Gamma \left(1+\frac{\sqrt{9 \gamma _{\mathit{eff}}^2-84 \gamma _{\mathit{eff}}+164}}
		{6 \gamma _{\mathit{eff}}-4}\right)\\
		&\times J_{\frac{\sqrt{9 \gamma _{\mathit{eff}}^2-84 \gamma _{\mathit{eff}}+164}}{6
				\gamma _{\mathit{eff}}-4}}\left(\frac{2 \Omega  \left(1+\frac{3}{2} \gamma _{\mathit{eff}} H_0 \left(t-t_0\right)\right){}^{1-\frac{2}{3 \gamma _{\mathit{eff}}}}}
		{ a_0 H_0\left(3 \gamma _{\mathit{eff}}-2\right)}\right)+C_6 \left(\frac{\Omega }
		{a_0 H_0 \left(3 \gamma _{\mathit{eff}}-2\right)}\right){}^{\frac{\sqrt{9 \gamma _{\mathit{eff}}^2-84
					\gamma _{\mathit{eff}}+164}}{6 \gamma _{\mathit{eff}}-4}}\\
		& \times \Gamma \left(1-\frac{\sqrt{9 \gamma _{\mathit{eff}}^2-84 \gamma _{\mathit{eff}}+164}}
		{6 \gamma _{\mathit{eff}}-4}\right)J_{-\frac{\sqrt{9 \gamma _{\mathit{eff}}^2-84
					\gamma _{\mathit{eff}}+164}}{6 \gamma _{\mathit{eff}}-4}}\left(\frac{2 \Omega
			\left(1+\frac{3}{2} \gamma _{\mathit{eff}} H_0 \left(t-t_0\right)\right){}^{1-\frac{2}{3
					\gamma _{\mathit{eff}}}}}{ a_0 H_0\left(3 \gamma _{\mathit{eff}}-2\right)}\right)\bigg]\,,
	\end{array}
\end{align}
where $C_{5}$ and $C_{6}$ are arbitrary constants. Now we study the evolution of GWs for different choices of EoS parameters.
In FIG. \ref{fig5}, the evolution of the waveform is depicted in a dust-dominated universe. Obviously, this figure shows a downward trend for three different model parameters: $\lambda$.
Note that $\lambda$ is restricted by observations as $-0.1<\lambda<1.5$ \cite{Velten:2017}. 

%%%%%%%%%%%%%%%%%%%%%%%%%%%%%%%%%%%%%%%%%%%%%%%%%%%%%%%%%%%%%%%%%%%
In this model, by comparing figures \ref{fig5} and \ref{fig7} with the behaviors of GWs in the dust-dominated and radiation-dominated scenarios of the GR theory, a significant similarity and consistency can be seen, which was expected due to the given choice of the form of the function $f(R,T)$. However, comparing FIG. \ref{fig9} with the results of the GR theory, we witness a much higher rate of decrease of frequency and amplitude of GWs in the $f(R,T)$ theory. This can be a response to the existence of a hypothetical form of dark energy in such theories, which leads to an increase in the background expansion rate. As the contribution of trace of the energy-momentum tensor (look at FIG. \ref{fig9}) in the action of this theory increases, the stability of the GWs diminishes.

As one can see from figures \ref{fig7} and \ref{fig8}, for the radiation-dominated universe of this model, there are more stable waves compared to the two former phases. The reason is that in the radiation-dominated universe, the expansion rate is slower than in the dust-dominated and quintessence phase.

%%%%%%%%%%%%%%%%%%%%%%%%%%%%%%%%%%%%%%%%%%%%%%%%%%%%%%%%%%%%%%%%%%%
% It is important to note that in all plots
%in this paper we assume $\kappa^{2}=a_{0}=1$.

\begin{figure}[H]
	\centering
	\includegraphics[scale=0.54]{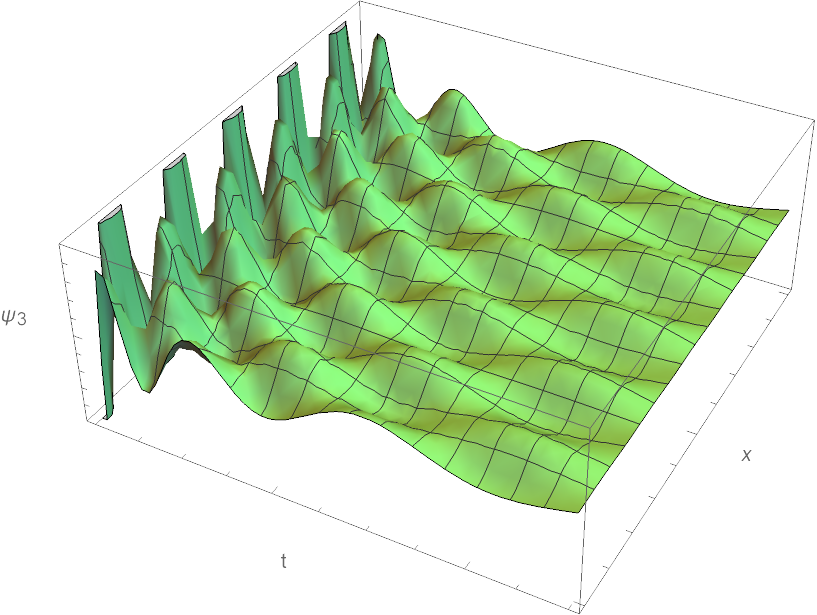}
	\caption{3D plot for the behavior of a test wave $\psi_{3}(x,t)=\rchi_{3}(x)\mathcal{T}_{3}(t)$
		for $f(R,T)$ gravity theory in dust-dominated flat FRW universe (t and x axes are arbitrary units).}
	\label{fig6}
\end{figure}

%%%%%%%%%%%%%%%%%%%%%%%%%%%%%%%%%%%%%

\begin{figure}[H]
	\centering
	\includegraphics[scale=0.45]{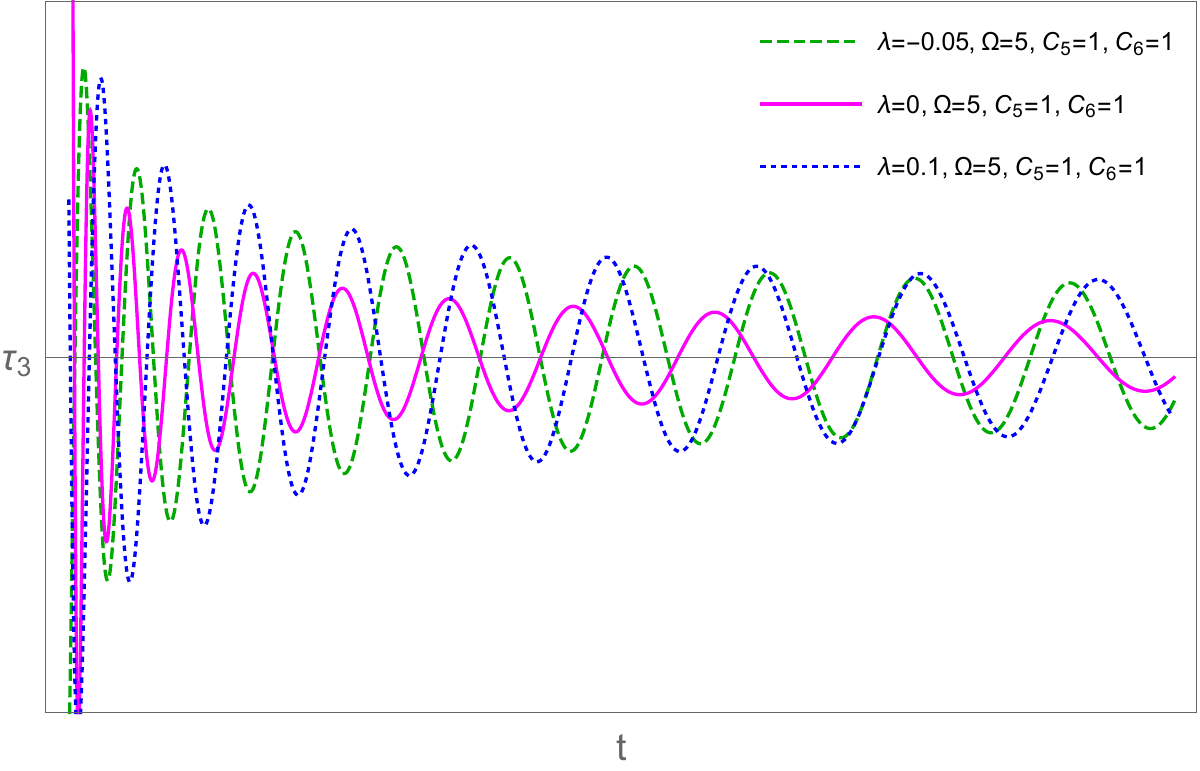}
	\caption{Evolution of gravitational wave for $f(R,T)$ gravity theory in radiation-dominated
		flat FRW universe ($\gamma=\frac{4}{3}$). (The time axis is an arbitrary unit)}
	\label{fig7}
\end{figure}

\begin{figure}[H]
	\centering
	\includegraphics[scale=0.61]{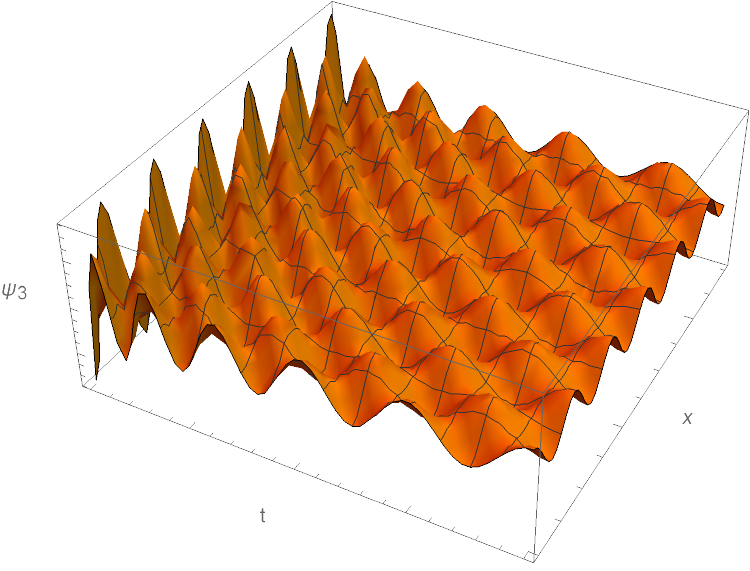}
	\caption{3D plot for the behavior of a test wave $\psi_{3}(x,t)=\rchi_{3}(x)\mathcal{T}_{3}(t)$
		for $f(R,T)$ gravity theory in radiation-dominated flat FRW universe ($\gamma=\frac{4}{3}$) (t and x axes are arbitrary units).}
	\label{fig8}
\end{figure}

%%%%%%%%%%%%%%%%%%%%%%%%%%%%%%%%%%%%%%%%%%%%%%%%%%%%%%%%%%%%%%%%%%%

\begin{figure}[H]
	\centering
	\includegraphics[scale=0.51]{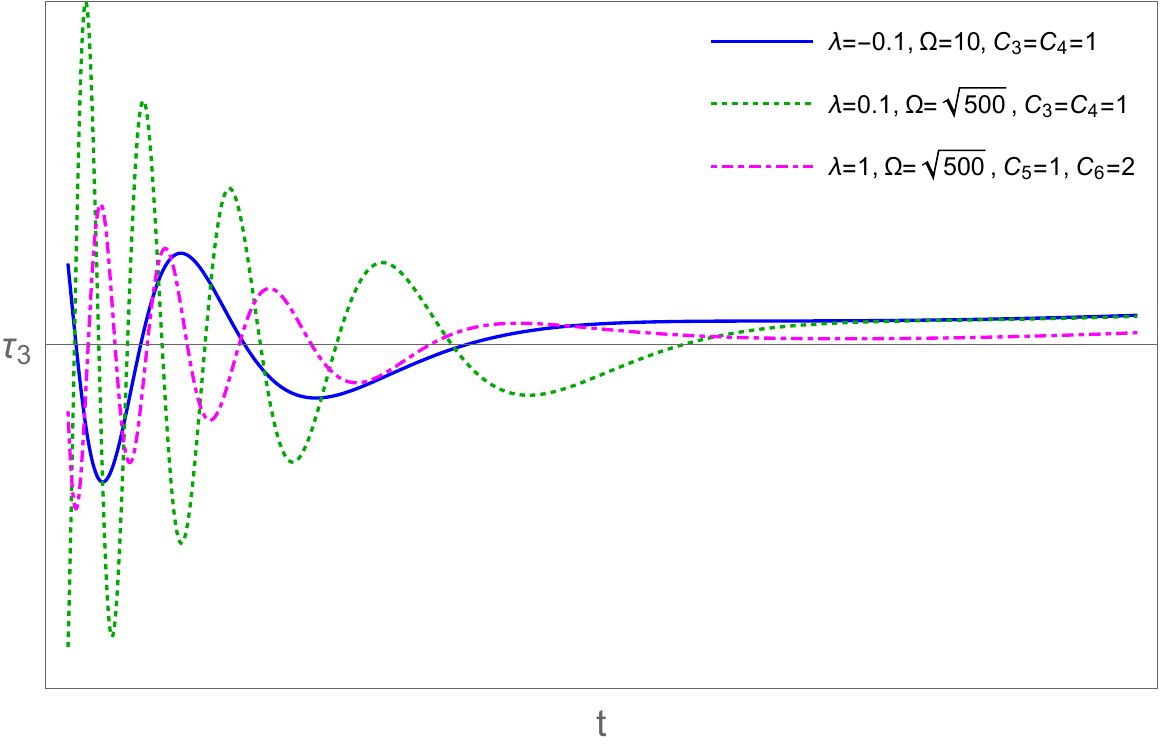}
	\caption{Evolution of gravitational wave for $f(R,T)$ gravity theory in quintessence regime
		FRW universe ($\gamma=\frac{1}{2}$). (The time axis is an arbitrary unit)}
	\label{fig9}
\end{figure}

\begin{figure}[H]
	\centering
	\includegraphics[scale=0.65]{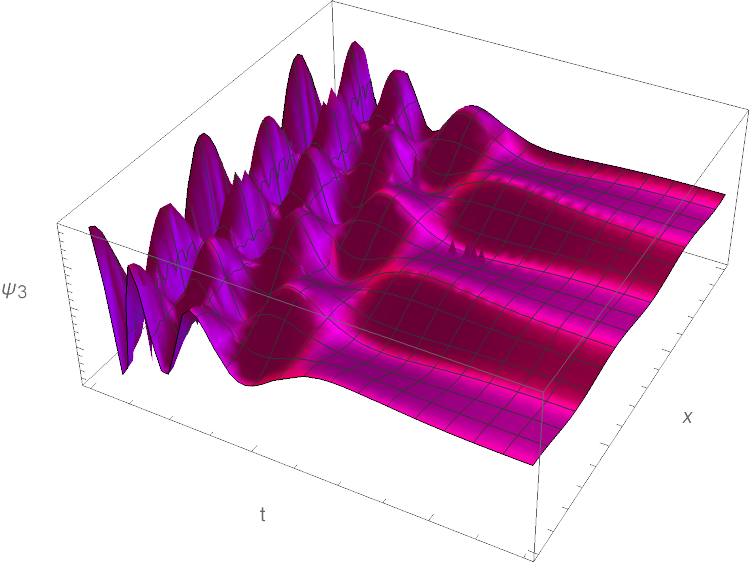}
	\caption{3D plot for the behavior of a test wave $\psi_{3}(x,t)=\rchi_{3}(x)\mathcal{T}_{3}(t)$
		for $f(R,T)$ gravity theory in quintessence regime FRW universe (t and x axes are arbitrary units).}
	\label{fig10}
\end{figure}

%%%%%%%%%%%%%%%%%%%%%%%%%%%%%%%%%%

\subsection{Evolution of the Waveform in \boldmath{$f(R,T_{\mu\nu} T^{\mu\nu})$} Gravity}
\begin{figure}[H]
     \centering
         \includegraphics[scale=0.59]{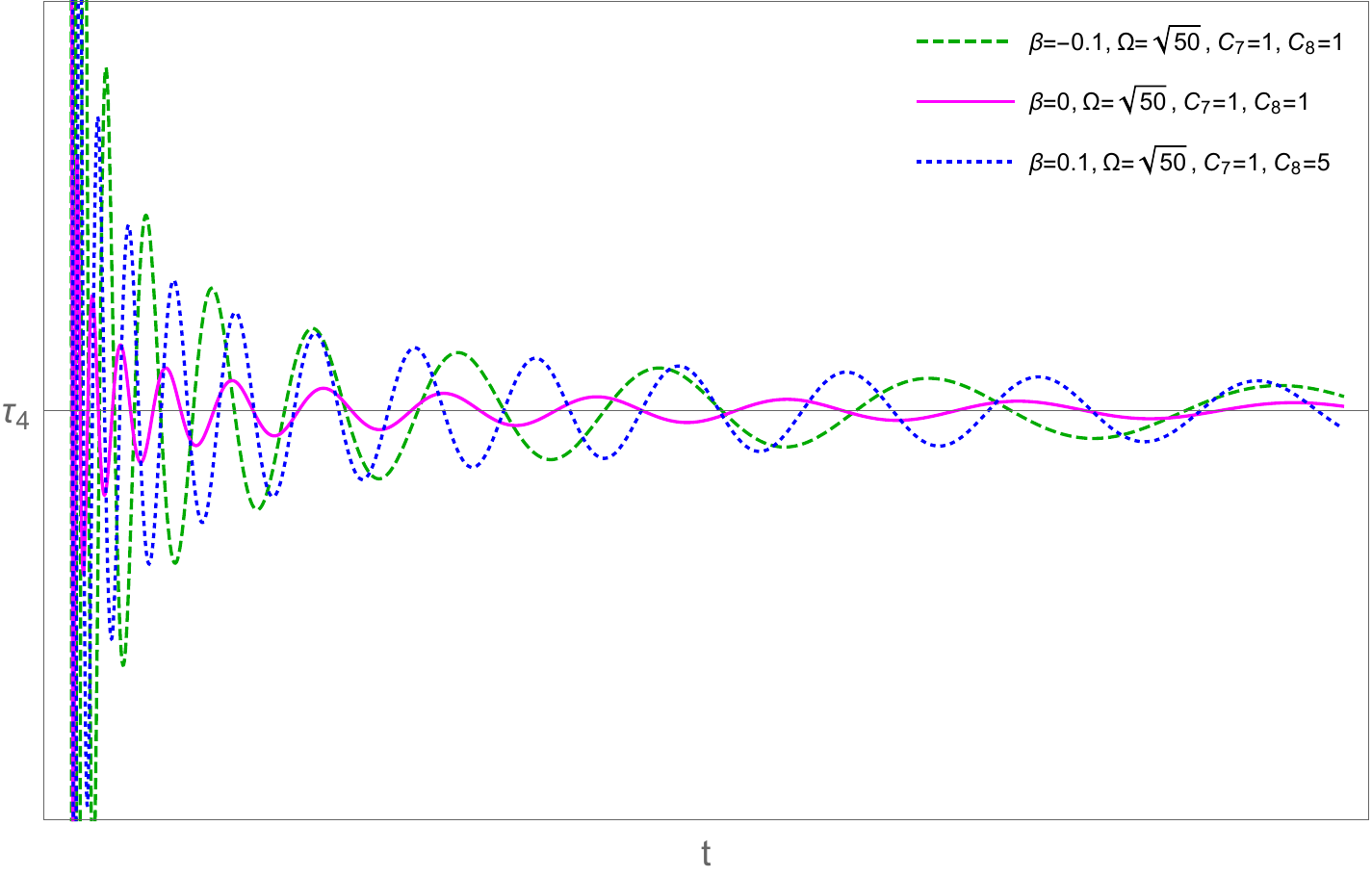}
         \caption{Evolution of gravitational wave for $f(R,T_{\mu\nu}T^{\mu\nu})$ gravity theory
         in dust-dominated flat FRW universe ($\gamma=1$). (The time axis is an arbitrary unit)}
         \label{fig11}
\end{figure}

To get a more clear image of the waveform for the considered model of $f(R,T_{\mu\nu} T^{\mu\nu})=\frac{R}{2\kappa^{2} }+\beta\sqrt{T_{\mu\nu}T^{\mu\nu}}$,
we use Eqs. (\ref{temporalpart}) and (\ref{scalefactorfrtt}) in a similar way, which yields
\begin{equation}
\frac{d^{2}\mathcal{T}_{4}}{d t^{2}}+\frac{2}{\gamma_{\mathit{eff}}\ t}\frac{d\mathcal{T}_{4}}{dt}+\left[\frac{\Omega^{2}}{t^{\frac{4}{3 \gamma_{\mathit{eff}}}}}  +\frac{4 \left(3 \gamma _{\mathit{eff}}-8\right)}{9\ \gamma _{\mathit{eff}}^2\ t^2}\right]\mathcal{T}_{4}=0.
\end{equation}
The above equation has a solution in the following form
\begin{align}
\begin{array}{ll}
\mathcal{T}_{4}(t) & = t^{\frac{1}{2}-\frac{1}{\gamma _{\mathit{eff}}}} \left(\frac{ \left(6 \gamma _{\mathit{eff}}-4\right)}{3 \Omega \gamma _{\mathit{eff}}}\right){}^{\frac{2}{3 \gamma _{\mathit{eff}}-2}-\frac{1}{2}}\bigg[C_7 \ \Gamma \left(1-\frac{\sqrt{9 \gamma _{\mathit{eff}}^2-84 \gamma _{\mathit{eff}}+164}}{6 \gamma _{\mathit{eff}}-4}\right) J_{-\frac{\sqrt{9 \gamma _{\mathit{eff}}^2-84 \gamma _{\mathit{eff}}+164}}{6 \gamma _{\mathit{eff}}-4}}\left(\frac{3 \Omega  \gamma _{\mathit{eff}} t^{1-\frac{2}{3 \gamma _{\mathit{eff}}}}}{ \left(3 \gamma _{\mathit{eff}}-2\right)}\right)\\
& + C_8 \ \Gamma \left(\frac{\sqrt{9 \gamma _{\mathit{eff}}^2-84 \gamma _{\mathit{eff}}+164}}{6 \gamma _{\mathit{eff}}-4}+1\right) J_{\frac{\sqrt{9 \gamma _{\mathit{eff}}^2-84 \gamma _{\mathit{eff}}+164}}{6 \gamma _{\mathit{eff}}-4}}\left(\frac{3 \Omega  \gamma _{\mathit{eff}} t^{1-\frac{2}{3 \gamma _{\mathit{eff}}}}}{ \left(3 \gamma _{\mathit{eff}}-2\right)}\right)\bigg]
\end{array}
\end{align}
where $C_7$ and $C_8$ are constants of integration.
%%%%%%%%%%%%%%%%%%%%%%%%%%%%%%%%%%%%%%%%%%%%%%%%%%%%%%%%%%%%%%
In this model, in the dust-dominated regime (FIG. \ref{fig11}), as the coupling parameter ($\beta$) increases, we observe an increase in the amplitude and frequency of GWs compared to the GR model, and the waves are more stable in this model. This maybe a result of increasing or decreasing in the rate of energy density reduction, which directly depends on the sign and magnitude of the parameter $\beta$. In this particular model, a positive sign ($\beta>0$) leads to an increase in the rate of energy density reduction, while a negative sign ($\beta<0$) leads to a decrease in the rate of matter dilution. However, an important aspect of this model relates to the radiation-dominated regime. Despite previous studies (see \cite{Katirci:2014}) predicting similar behavior in terms of the dependence of the rate of energy density reduction on the magnitude and sign of the parameter $\beta$, an increase in $\beta$, the amplitude of GWs decay compared to the GR model, but the stability of GWs does not change remarkably.

Regarding the quintessence regime, the behavior of GWs (faster decay compared to the GR model) is entirely expected due to the dark energy density, which leads to a background expansion.

%%%%%%%%%%%%%%%%%%%%%%%%%%%%%%%%%%%%%%%%%%%%%%%%%%%%%%%%%%%%%%%%%%%

\begin{figure}[H]
         \centering
         \includegraphics[scale=0.60]{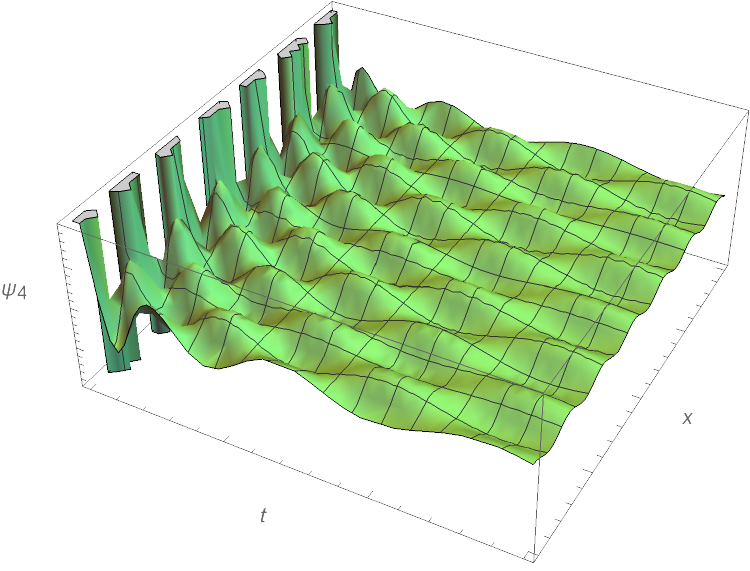}
         \caption{3D plot for the behavior of a test wave $\psi_{4}(x,t)=\rchi_{4}(x)\mathcal{T}_{4}(t)$
         for $f(R,T_{\mu\nu}T^{\mu\nu})$ gravity theory in dust-dominated flat FRW universe (t and x axes are arbitrary unit).}
         \label{fig12}
\end{figure}

%%%%%%%%%%%%%%%%%%%%%%%%%%%%%%%%%%%%%%%%%%%%%%%%%%%%%%%%%%%%%%%%%%

\begin{figure}[H]
     \centering
         \includegraphics[scale=0.55]{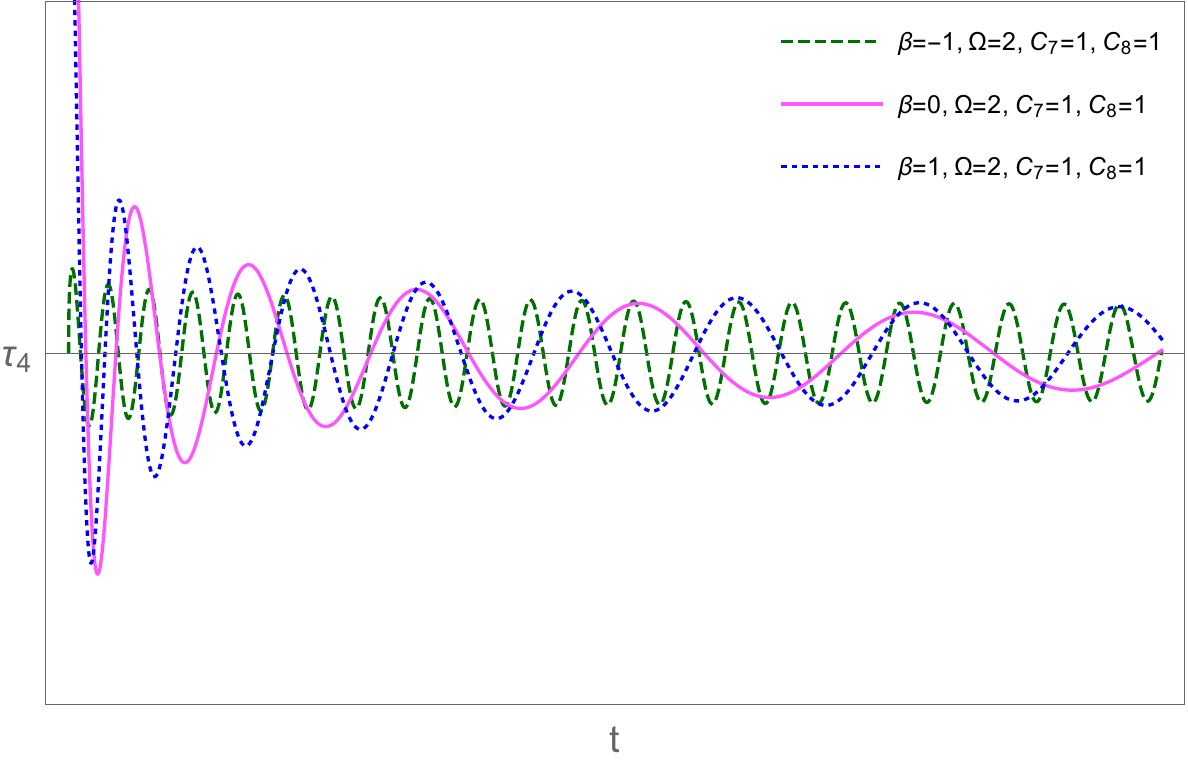}
         \caption{Evolution of gravitational wave for $f(R,T_{\mu\nu}T^{\mu\nu})$ gravity theory
         in radiation-dominated flat FRW universe ($\gamma=\frac{4}{3}$). (The time axis is an arbitrary unit)}
         \label{fig13}
\end{figure}

\begin{figure}[H]
         \centering
         \includegraphics[scale=0.56]{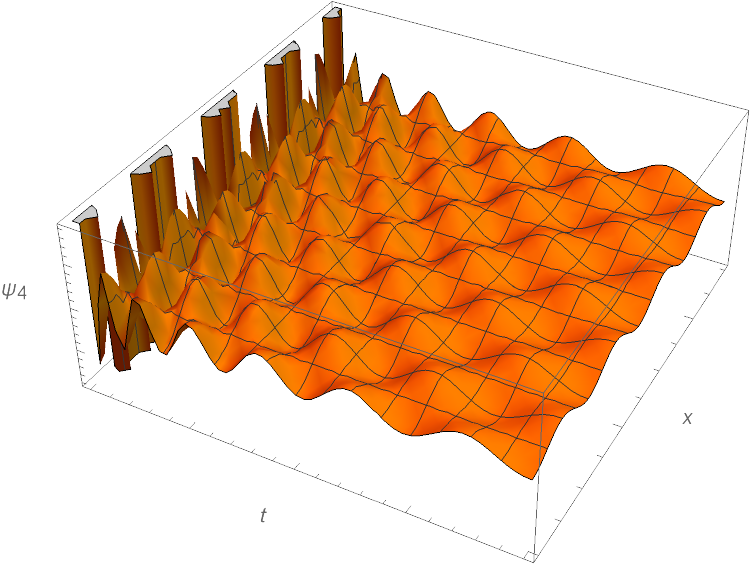}
         \caption{3D plot for the behavior of a test wave $\psi_{4}(x,t)=\rchi_{4}(x)\mathcal{T}_{4}(t)$
         for $f(R,T_{\mu\nu}T^{\mu\nu})$ gravity theory in radiation- dominated flat FRW universe (t and x axes are arbitrary unit).}
         \label{fig14}
\end{figure}

%%%%%%%%%%%%%%%%%%%%%%%%%%%%%%%%%%%%%
\begin{figure}[H]
     \centering
         \includegraphics[scale=0.55]{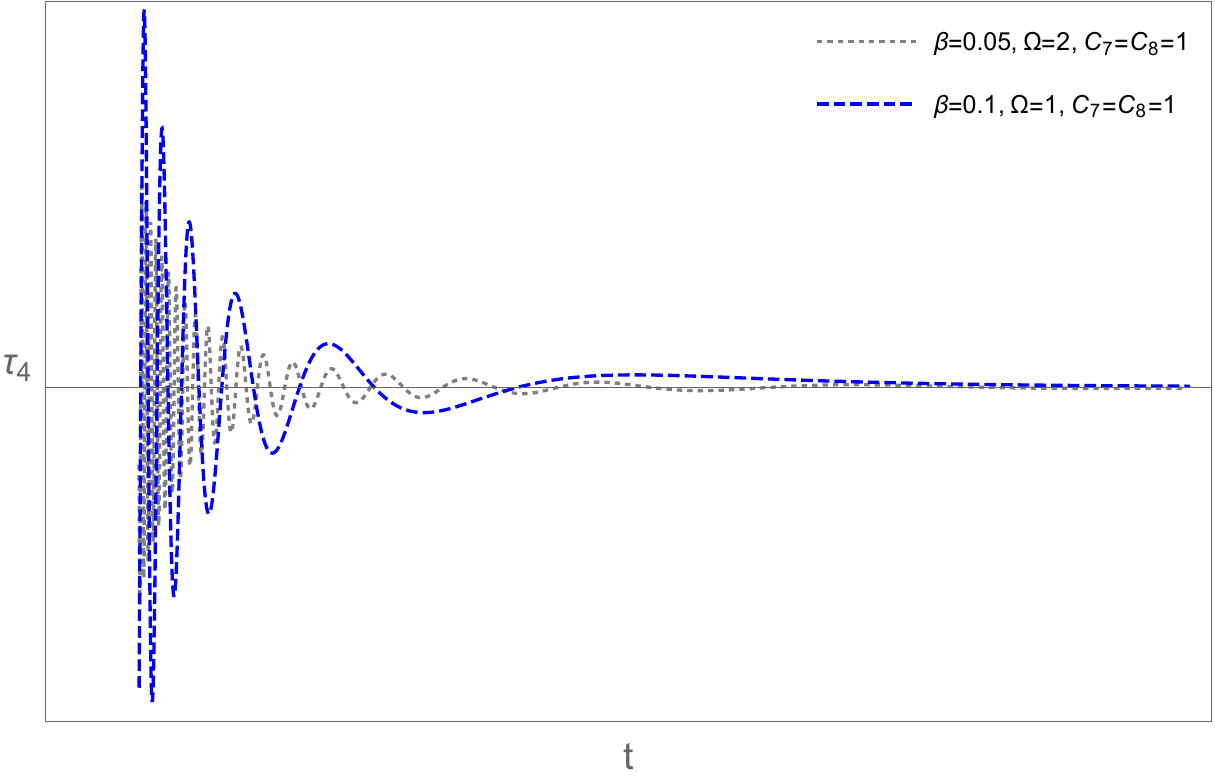}
         \caption{Evolution of gravitational wave for $f(R,T_{\mu\nu}T^{\mu\nu})$ gravity theory
         in quintessence regime flat FRW universe ($\gamma=\frac{1}{3}$). (The time axis is an arbitrary unit)}
         \label{fig15}
\end{figure}

\begin{figure}[H]
         \centering
         \includegraphics[scale=0.60]{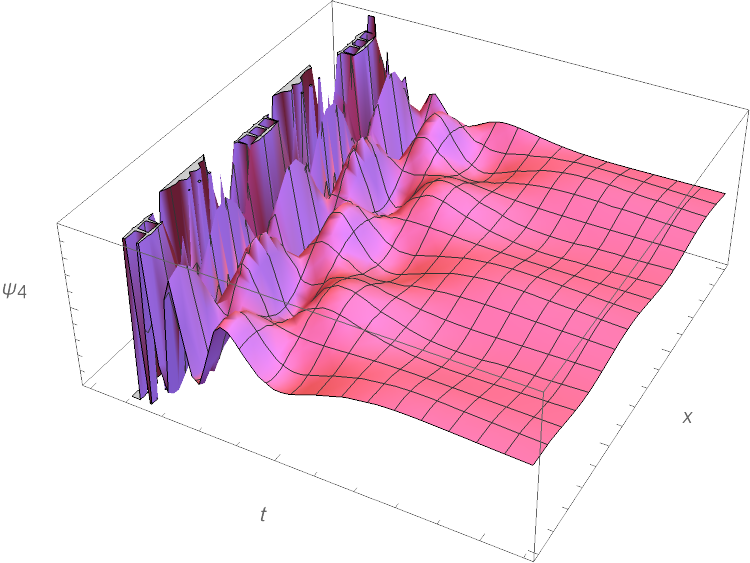}
         \caption{3D plot for the behavior of a test wave $\psi_{4}(x,t)=\rchi_{4}(x)\mathcal{T}_{4}(t)$
         for $f(R,T_{\mu\nu}T^{\mu\nu})$ gravity theory in quintessence regime flat FRW universe (t and x axes are arbitrary unit).}
         \label{fig16}
\end{figure}

\subsection{Evolution of the Waveform in \boldmath{$f(Q,T)$} Gravity}
\begin{figure}[H]
	\centering
	\includegraphics[scale=0.55]{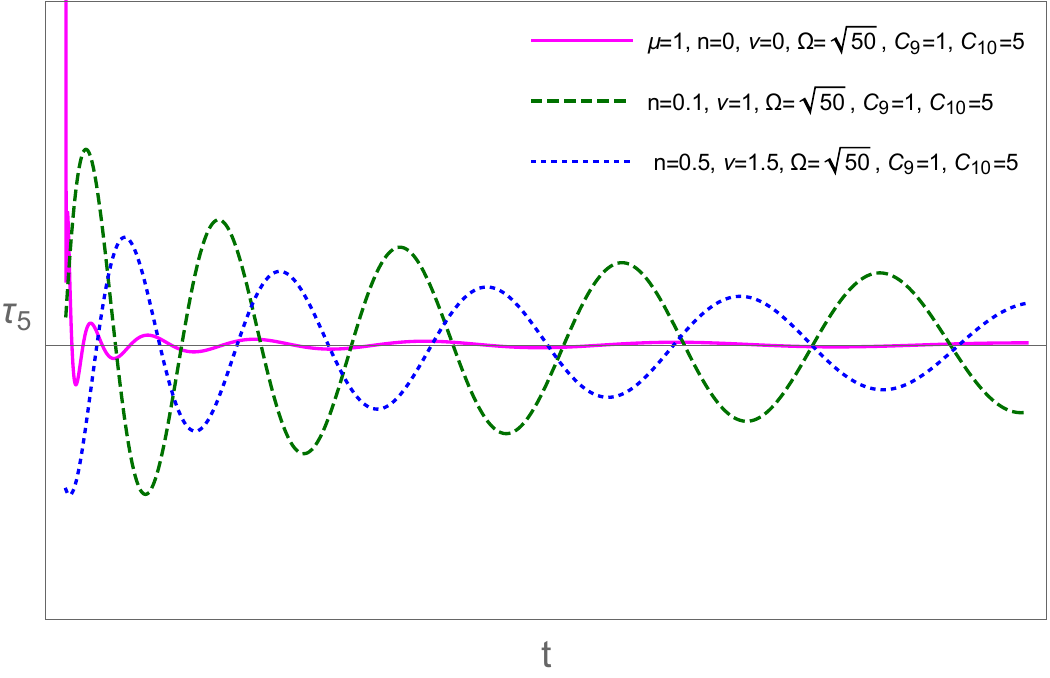}
	\caption{Evolution of gravitational wave for $f(Q,T)$ gravity theory in dust-dominated flat
		FRW universe($\gamma=1$). (The time axis is an arbitrary unit)}
	\label{fig17}
\end{figure}

\begin{figure}[H]
	\centering
	\includegraphics[scale=0.48]{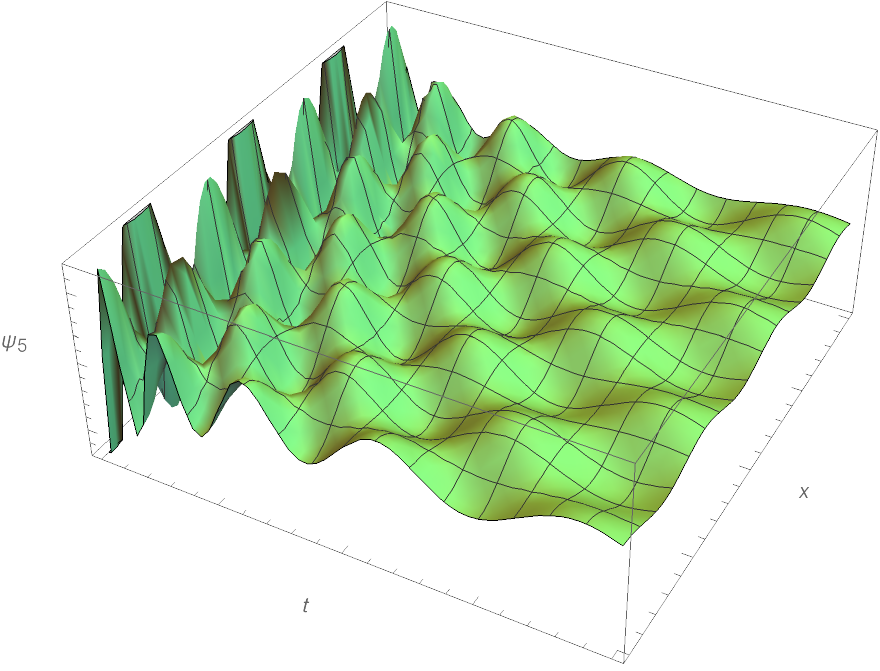}
	\caption{3D plot for the behavior of a test wave $\psi_{5}(x,t)=\rchi_{5}(x)\mathcal{T}_{5}(t)$ for
		$f(Q,T)$ gravity theory in dust-dominated flat FRW universe (t and x axes are arbitrary units).}
	\label{fig18}
\end{figure}

\begin{figure}[H]
	\centering
	\includegraphics[scale=0.46]{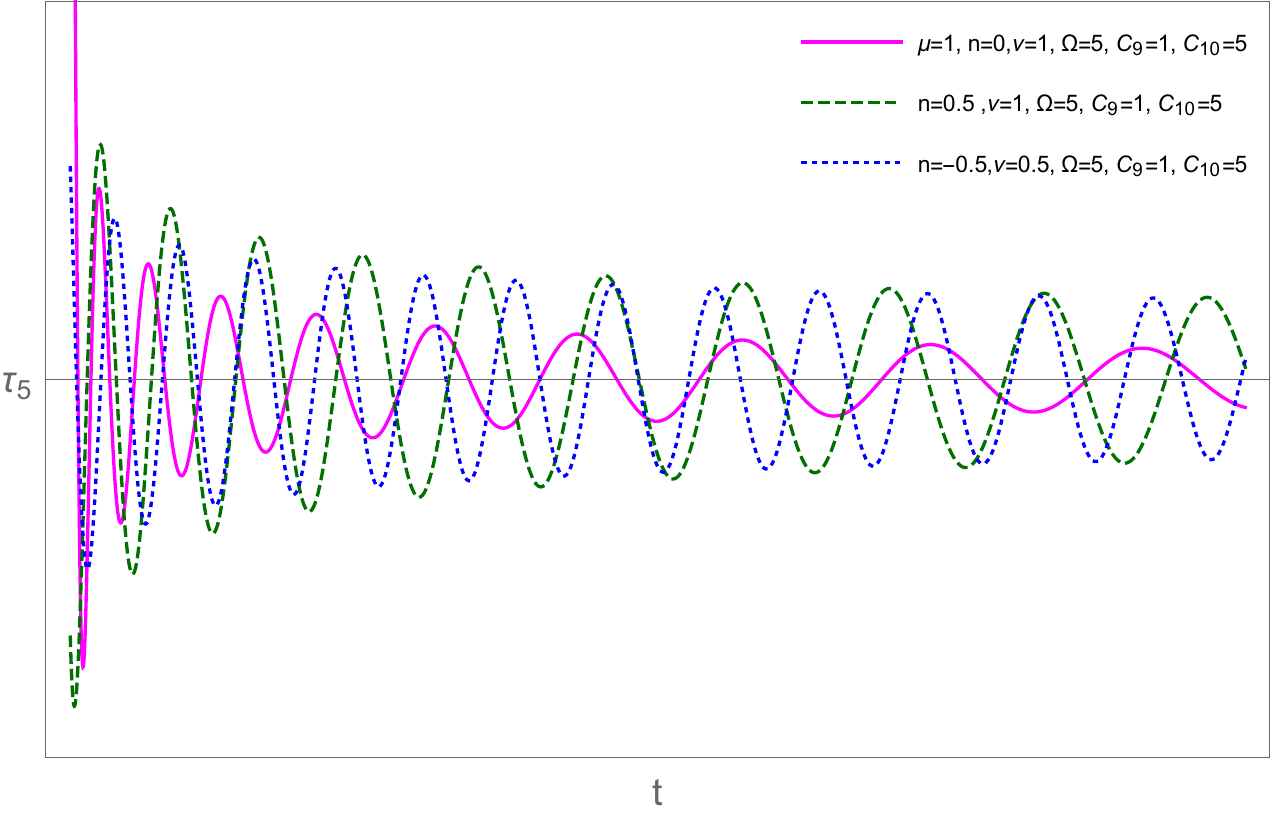}
	\caption{Evolution of gravitational wave for $f(Q,T)$ gravity theory in radiation-dominated
		flat FRW universe ($\gamma=\frac{4}{3}$). (The time axis is an arbitrary unit)}
	\label{fig19}
\end{figure}

\begin{figure}[H]
	\centering
	\includegraphics[scale=0.58]{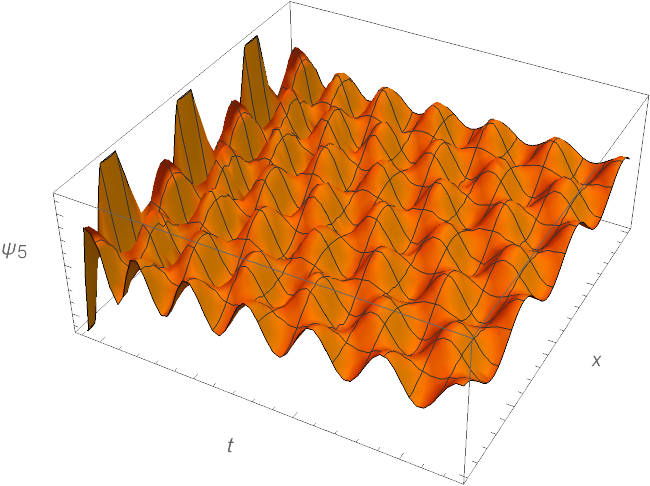}
	\caption{3D plot for the behavior of a test wave $\psi_{5}(x,t)=\rchi_{5}(x)\mathcal{T}_{5}(t)$
		for $f(Q,T)$ gravity theory in radiation-dominated flat FRW universe (t and x axes are arbitrary units).}
	\label{fig20}
\end{figure}

\begin{figure}[H]
	\centering
	\includegraphics[scale=0.5]{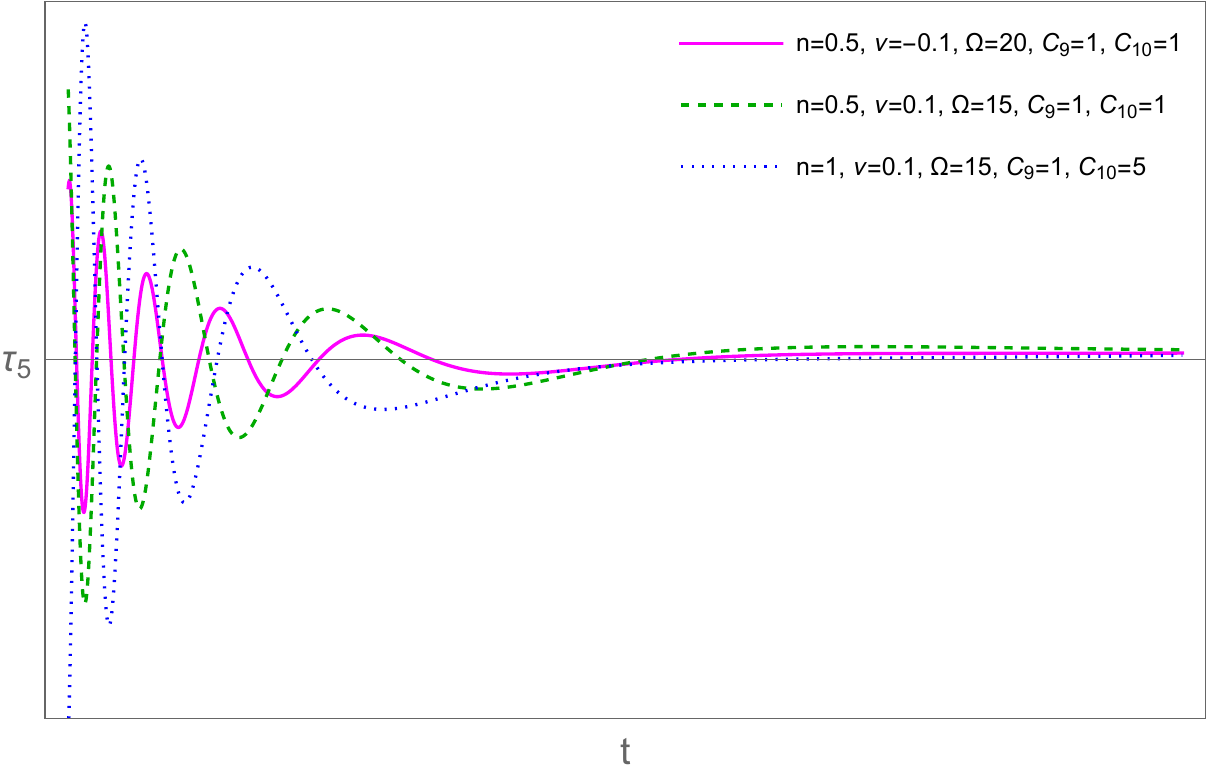}
	\caption{Evolution of gravitational wave for $f(Q,T)$ gravity theory in quintessence regime
		flat FRW universe($\gamma=\frac{1}{2}$). (The time axis is an arbitrary unit)}
	\label{fig21}
\end{figure}

\begin{figure}[H]
	\centering
	\includegraphics[scale=0.6]{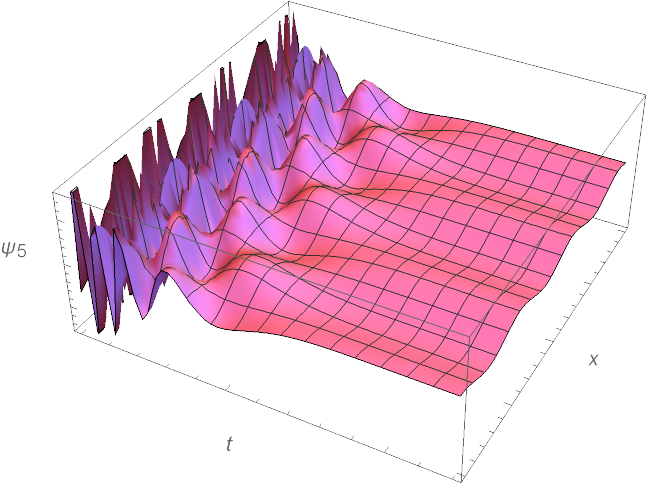}
	\caption{3D plot for the behavior of a test wave $\psi_{5}(x,t)=\rchi_{5}(x)\mathcal{T}_{5}(t)$
		for $f(Q,T)$ gravity theory in quintessence regime flat FRW universe (t and x axes are arbitrary units).}
	\label{fig22}
\end{figure}

According to (\ref{scffqt}), the temporal evolution of the wave of this type of gravity is
given by the following differential equation. By putting the scale factor (\ref{scffqt}) into
(\ref{scalefactorfrt}), the temporal evolution of the wave in this case provides
\begin{equation}
$$\scalemath{0.83}{\frac{d^{2}\mathcal{T}_{5}}{dt^{2}}+\frac{3a_{1} H_0}{a_{1}+3\gamma (\kappa^{2}+\nu ) H_0 \left(t-t_0\right)} \frac{d\mathcal{T}_{5}}{dt}+\left[\frac{\Omega ^2 }{a_0^2 \ \bigg(a_{1}+ 3 \gamma (\kappa^{2}+\nu ) H_0 \left(t-t_0\right)\bigg){}^{\frac{2a_{1}}{3 \gamma(\kappa^{2}+\nu ) }}}+\frac{2\ a_{1} H_0^2 \left(3 \gamma(\kappa^{2}+\nu ) -4 a_{1}\right)}{\bigg(a_{1}+ 3 \gamma (\kappa^{2}+\nu ) H_0 \left(t-t_0\right)\bigg){}^2}\right]\mathcal{T}_{5} =0}$$,
\end{equation}

where we have defined $a_{1}=(n+1) \left(2\kappa^{2} -\nu(\gamma -4)\right)$.
%\begin{align}
%	\begin{array}{ll}
%		\mathcal{T}_{5}(t) & = \left(\frac{a_1}{2}\left[2+3\zeta H_0 \left(t-t_0\right)\right] \right){}^
%		{\frac{1}{2}-\frac{1}{\zeta}} \bigg[C_{9}
%		\frac{\Gamma \left(1+\frac{\zeta^2\sqrt{9 \zeta^2-84 \zeta+164}}
%			{6 \zeta-4}\right)}{\left(\frac{\Omega }{a_0 a_1 H_0 \left(3 \zeta-2\right)}\right){}^
%			{\frac{\zeta^2\sqrt{9 \zeta^2-84 \zeta+164}}{6 \zeta-4}}}\\
%		& J_{\frac{\zeta^2\sqrt{9 \zeta^2-84 \zeta+164}}{6 \zeta-4}}\left(\frac{ \Omega  \bigg(a_1(2+3 \zeta
%			H_0 \left(t-t_0\right))\bigg){}^{1-\frac{2}{3 \zeta}}}
%		{ a_0 a_1 H_0\left(3 \zeta-2\right)}\right)\\
%		&+C_{10} \frac{\Gamma \left(1-\frac{\zeta^2\sqrt{9 \zeta^2-84 \zeta+164}}
%			{6 \zeta-4}\right)}{\left(\frac{\Omega }{a_0 a_1 H_0 \left(3 \zeta-2\right)}\right){}^
%			{-\frac{\zeta^2\sqrt{9 \zeta^2-84 \zeta+164}}{6 \zeta-4}}}J_{-\frac{\zeta^2\sqrt{9 \zeta^2-84
%					\zeta+164}}{6 \zeta-4}}\left(\frac{\Omega
%			\bigg(a_1\left(2+3 \zeta H_0 (t-t_0)\right)\bigg){}^{1-\frac{2}{3
%					\zeta}}}{ a_0 a_1 H_0\left(3 \zeta-2\right)}\right)\bigg]
%	\end{array}
%\end{align}
The solution of the above equation is obtained as follows

\begin{align}
	\begin{array}{ll}
		$$\scalemath{0.9}{\mathcal{T}_{5}(t)}$$&=$$\scalemath{0.85}{ \left(\frac{a_1}{2}\left[2+3\zeta H_0 \left(t-t_0\right)\right] \right){}^
		{\frac{1}{2}-\frac{1}{\zeta}} \bigg[C_{9}
		\frac{\Gamma \left(1+\frac{\zeta^2\sqrt{9 \zeta^2-84 \zeta+164}}
			{6 \zeta-4}\right)}{\left(\frac{\Omega }{a_0 a_1 H_0 \left(3 \zeta-2\right)}\right){}^
			{\frac{\zeta^2\sqrt{9 \zeta^2-84 \zeta+164}}{6 \zeta-4}}} J_{\frac{\zeta^2\sqrt{9 \zeta^2-84 \zeta+164}}{6 \zeta-4}}\left(\frac{ \Omega  \bigg(a_1(2+3 \zeta
			H_0 \left(t-t_0\right))\bigg){}^{1-\frac{2}{3 \zeta}}}{ a_0 a_1 H_0\left(3 \zeta-2\right)}\right)}$$\\
		&$$\scalemath{0.85}{+C_{10} \frac{\Gamma \left(1-\frac{\zeta^2\sqrt{9 \zeta^2-84 \zeta+164}}
			{6 \zeta-4}\right)}{\left(\frac{\Omega }{a_0 a_1 H_0 \left(3 \zeta-2\right)}\right){}^
			{-\frac{\zeta^2\sqrt{9 \zeta^2-84 \zeta+164}}{6 \zeta-4}}}J_{-\frac{\zeta^2\sqrt{9 \zeta^2-84
					\zeta+164}}{6 \zeta-4}}\left(\frac{\Omega
			\bigg(a_1\left(2+3 \zeta H_0 (t-t_0)\right)\bigg){}^{1-\frac{2}{3
					\zeta}}}{ a_0 a_1 H_0\left(3 \zeta-2\right)}\right)\bigg]}$$,
	\end{array}
\end{align}

where $C_{9}$ and $C_{10}$ are constants of integration and we have defined $\zeta= 2\gamma(\kappa ^2+ \nu) /a_{1}$. We plot the waveforms
for this model in figures \ref{fig17}-\ref{fig22}. 
%%%%%%%%%%%%%%%%%%%%%%%%%%%%%%%%%%%
In this model, by comparing the gravitational wave behavior with similar results in GR, we observe a striking similarities in both dust-dominated and radiation-dominated epoches. It is demonstrated that despite its ability to explain the accelerated expansion of the universe without the need for dark energy, this theory exhibits many similarities with the GR model in terms of GWs behavior. Waves from the quintessence regime have undertaken an initial marked fall, which is quite expected because of the negative pressure produced by the dark energy (see figures  \ref{fig21}  and \ref{fig22} ).

%%%%%%%%%%%%%%%%%%%%%%%%%%%%%%%%%%%%%%%%%%%%%%%%%%%%%%

%%%%%%%%%%%%%%%%%%%%%%%%%%%%%%%

\section{Conclusions}\label{conclusion}
One interesting recent approach to describe the universe's accelerating phase is to introduce a non-minimal coupling between matter and geometry in the Einstein-Hilbert action. In principle, many types of such coupling are possible, including the non-minimally coupled scalar theory, coupling
between the Ricci scalar and the trace of the matter energy-momentum tensor, a non-minimally coupled torsion-matter theory, coupling between the Ricci scalar and the norm of the energy-momentum tensor, and so on.
In this paper, we have studied the implications of some specific curvature-matter coupling models on the gravitational waves traveling in a cosmological background through the cosmic fluid.
As an initial step towards studying GW's propagation in these models, we have obtained the form of the wave equation in flat FRW space-time. Secondly, we have investigated the background scale factor evolution of the non-minimal coupling gravity models to substitute them within (the temporal part of) the wave equation that we obtained earlier and get an exact solution for the temporal and spatial parts.
Investigating the obtained solution of the wave equation for the considered models with different matter contributions, and compare them with the results obtained from GR studies in different cosmological backgrounds \cite{Mondal:2021}, we have found that in most cases, GWs' behavior is almost the same, except in presence of dark matter, which both the amplitude and frequency of the GWs decay faster with time. In the first model, a non-minimally coupled scalar theory, we have two scale factors for two cases: $\alpha>2$ and $\alpha<-1$ corresponding to the quintessence phase and phantom phase, respectively. The comparison of our proposed non-minimal coupling theory for $\alpha>2$ with GR highlights the similarities with the dust-dominated case. However, the rate of reduction is higher in the non-minimal coupling theory due to the higher rate of cosmic expansion associated with this model. In the case of $\alpha<-1$, GWs amplitude and frequency decay more remarkably as an effective phantom phase triggers perpetual expansion in the universe. The similar evolution of GWs in this model and GR suggests that the selection of appropriate values for $\alpha$ and constant can offer insights into the behavior of GWs in general.
%We have shown that the rate of decrease in amplitude of the wave is significantly faster for $\alpha<-1$. In other words, for more negative powers of the curvature, the amplitude and frequency of the wave decrease more slowly. 
Further, for $f(R,T)$ gravity, by considering a simple linear model, a scale factor is obtained in terms of the model parameters and EoS parameter, as well. Hence, we can study the propagation of GWs in different phases of the evolution of the universe, particularly in the dust-dominated, radiation-dominated, and quintessence regimes. The analysis has shown similar behavior of GWs in dust-dominated and radiation-dominated to GR which is a result of the given linear form of $f(R,T)$. However, when it comes to quintessence phase, GWs decay significantly and are less stable as a response to higher speed of expansion due to the presence of dark matter.

%for the radiation-dominated case, despite decaying in both amplitude and frequency, the waves are more stable in comparison to the other two phases due to the slower speed of expansion.

In addition, we employed a specific case of energy-momentum-powered gravity to analyze what impression the background scale factor makes on the waveforms. Our findings demonstrate that the behavior of GWs is influenced by the specific regime under consideration. The dust-dominated regime exhibited amplified amplitudes and higher frequencies compared to GR with increased stability as $\beta$ increased. The radiation-dominated regime, contrary to previous expectations, displayed a decay in amplitude with no substantial impact on stability. The quintessence regime showed faster decay in GWs due to the influence of dark energy density on the background expansion.
%We have shown that gravitational waves exhibit different decreasing behaviors in different cosmic fluids. For example, for $\gamma=1/3$, the negative pressure of dark energy triggers the dramatic decaying of the amplitude and frequency of waves at the beginning of time. As a result, we cannot trace them in the present time. On the other hand, the waves from the dust-dominated and radiation-dominated eras can be detected in the considered model.

Moreover, we have assumed a functional form for $f(Q,T)$ that has an exact solution for the background scale factor, which could help us to look further into the fate of the GWs produced in different universe distributions. We have observed remarkable similarities in the behavior of GWs during both the dust-dominated and radiation-dominated epochs in $f(Q,T)$ model with GR model. 

Finally, it needs to be mentioned that the study of behavior of the gravitational waves, specifically the dynamics of their amplitude and frequency,in different modified gravity theories and comparing them to diverse observational data sets may  serve as an effective tool for constraining the different parameters of these theories.

\end{document}